\documentclass[12pt]{article}

\usepackage{latexsym}

\usepackage[font=small,labelfont=bf]{caption}

\usepackage{graphics}
\usepackage{graphicx}
\usepackage{epstopdf}
\usepackage{amssymb}
\usepackage{caption}
\usepackage{subcaption}
\usepackage{tabularx}
\usepackage{gensymb}
\usepackage{mathtools,amsmath,amssymb,mathrsfs,color,esint}
\usepackage{array}
\usepackage{makecell}
\usepackage{float}% to keep pictures in place
\usepackage{dcolumn}% Align table columns on decimal point
\usepackage{hyperref}
\usepackage{xcolor}

\begin{document}

\title{Polarized equatorial emission around regular compact objects}

\author{
Tsanimir Angelov$^{1}$\footnote{E-mail: \texttt{tsangelov@phys.uni-sofia.bg}}, \,
Rasim  Bekir$^{1}$\footnote{E-mail: \texttt{rbekir@uni-sofia.bg}}, \,
Galin Gyulchev$^{1}$\footnote{E-mail: \texttt{gyulchev@phys.uni-sofia.bg}},  \, \\
Petya Nedkova$^{1}$\footnote{E-mail: \texttt{pnedkova@phys.uni-sofia.bg}}, \,
Stoytcho Yazadjiev$^{1,2}$\footnote{E-mail: \texttt{yazad@phys.uni-sofia.bg}}\\ \\
 {\footnotesize${}^{1}$ Faculty of Physics, Sofia University ``St. Kliment Ohridski'',}\\
  {\footnotesize 5 James Bourchier Boulevard, Sofia~1164, Bulgaria } \\
    {\footnotesize${}^{2}$ Institute of Mathematics and Informatics,}\\
{\footnotesize Bulgarian Academy of Sciences, Academic G. Bonchev 8, } {\footnotesize Sofia 1113, Bulgaria}}
\date{}
\maketitle

\begin{abstract}
We study the polarization properties of regular compact objects motivated by quantum gravity. These geometries include static spherically symmetric non-singular black holes which undergo a continuous transition to regular particle-like spacetimes in certain regions of the parametric space. Applying a phenomenological model of a magnetized fluid ring orbiting in the equatorial plane, we search for characteristic signatures in the direct polarized images which could distinguish them from black holes in general relativity. At low inclinations the regular compact objects lead to a very similar polarization pattern as the Schwarzschild black hole. At high inclinations we observe a characteristic signature for the particle-like geometries. A second direct image of the fluid ring is formed which is absent for black holes. It may possess with three orders of magnitude higher polarization intensity and a specific polarization direction making it observationally significant. We elucidate the physical mechanism for the formation of the second direct image and pinpoint the plasma configurations which lead to amplification of its polarized emission.
\end{abstract}

\section{Introduction}

An intrinsic feature of general relativity is the presence of curvature singularities in black holes and cosmology. Assuming the weak energy condition and global hyperbolicity Penrose's singularity theorem associates the formation of a trapped region in gravitational collapse with the formation of a singularity in its interior \cite{Penrose:1965}- \cite{Hawking:1970}. It is generally believed that this behavior will be resolved in a viable theory of quantum gravity. Different theories modifying general relativity in the high curvature regime predict black holes which do not contain a curvature singularity. Such solutions called regular black holes arise for example in loop quantum gravity \cite{Perez:2017}-\cite{Rovelli:2017}, asymptotically safe quantum gravity \cite{Bonanno:2000}-\cite{Torres:2014}, and
noncommutative gravity \cite{Nicolini:2006}.

Regardless of the underlying fundamental theory regular black hole models can be developed phenomenologically by adopting the desired modifications at high curvature. It is generally assumed that there exist a certain critical energy leading to associated length and curvature scales. When the characteristic curvature scale is approached the spacetime metric describing a black hole in general relativity should be modified while the corrections remain minimal in the region of low curvature. The spacetime should further satisfy the limited curvature conjecture which states that the curvature invariants should be uniformly restricted by some universal value \cite{Markov:1982}-\cite{Polchinski:1989}.

Following these basic principles regular black holes were constructed which behave at large distances as the classical black holes in general relativity but the black hole interior is replaced by a regular de Sitter core. The first model of a non-singular black hole was developed by Bardeen \cite{Bardeen:1968} by considering the collapse of charged matter which leads to the formation of a charged matter core inside the black hole instead of a singularity. A closely related static and spherically symmetric metric was proposed by Hayward \cite{Hayward:2006} which was further generalized by Frolov and Zelnikov by exploring a more diverse class of mass functions \cite{Frolov:2016}-\cite{Frolov:2017}. Another line of research proposed a regularization procedure which leads to a black hole interior approaching the Minkowski spacetime at small distances \cite{Simpson:2019}. A variety of regular black hole solutions were constructed recently including also rotating spacetimes \cite{Dymnikova:1992}-\cite{Bonanno:2024} with topic developments summarized in the reviews \cite{Ansoldi:2008}-\cite{Lan:2023}.

Many of the regular black hole solutions allow for a continuous transition to regular horizonless compact objects. For a certain mass the transition is controlled by the critical length scale parameter introduced for regularizing the singularity. After a certain value of the length scale parameter the black hole horizon disintegrates and we obtain a regular particle-like solution. Such geometries are valuable laboratories for testing quantum gravity phenomenology. Since they lack a horizon phenomenological effects produced by the interior high-density core become theoretically observable and we get access to observational signatures carrying imprints of Plank-scale physics.

The aim of this paper is to study phenomenological effects caused by light propagation in the vicinity of horizonless compact objects. In particular we investigate the polarization of the electromagnetic radiation motivated by recent polarimetric observations of the galactic targets M87* and Sgr A* by the Event Horizon Telescope \cite{EHT1}-\cite{EHT8}. We consider both particle-like solutions with de Sitter cores such as the Bardeen, Hayward and Frolov spacetimes, and particle-like solutions containing Minkowski cores such as the Simpson-Viser geometry. These compact objects are characterized by two phenomenological regimes. For a certain range of the length scale parameter they represent ultracompact objects possessing a stable and an unstable light rings, while after a certain critical value they contain no light rings producing distinctive phenomenological behavior. The observational signatures of such particle-like solutions were investigated in a variety of works including studies of the gravitational lensing and the properties of the accretion disks in their vicinity \cite{Schee:2015}-\cite{Eichhorn:2024}.

In this work we study the polarization of the electromagnetic radiation emitted by inner regions of the accretion disks around particle-like compact objects. We adopt an analytical model in which the emitting region is approximated by a thin fluid ring orbiting in the equatorial plane around the central compact object. The fluid interacts with a uniformly magnetized medium and emits linearly polarized synchrotron radiation. This model was originally proposed by the Event Horizon Collaboration for describing the polarized radiation from the galactic target M87* \cite{Narayan:2021}-\cite{Gelles:2021}. Despite its simplicity it proved to reproduce successfully the main qualitative features of the observable polarization pattern of M87*. The phenomenological fluid ring model has been used for investigating the polarimetric properties of various compact objects of different physical nature in relation to the physical setting of the galactic targets M87* and Sgr A*. These studies include black holes in the modified theories of gravity \cite{Qin:2021}-\cite{Zhang:2022} as well as horizonless compact objects as  wormholes \cite{Delijski:2023}, naked singularities \cite{Nedkova:2023} and exotic stars \cite{Tamm:2025}.

The paper is organized as follows. In the next section we present the regular black hole geometries which we investigate and their particle-like regimes. In Section 3 we review the model of synchrotron emitting fluid orbiting in the equatorial plane focusing on the observable properties of the polarized radiation and the procedure for their computation. In Section 4 we describe a semi-analytical procedure for computing the images of a circular emitter. This procedure allow for the evaluation of the number of images of each lensing order which will be produced and their relative position on the observer's sky.  In Section 5 we present our simulations for the observable linear polarization for regular black holes and particle-like solutions at small inclinations. These studies are motivated by the observational settings of the galactic targets M87* and Sgr A* and reflect the capacity for differentiating between the Schwarzschild black hole and quantum gravity motivated compact objects by means of the Event Horizon Telescope. In Section 5 we simulate the observable polarization for particle-like solutions at high inclinations. In this regime the horizonless spacetimes lead to observable properties which are absent for black holes and may be used as a signature for their detection. In the last section we summarize our results.

\section{Regular black holes and particle-like geometries}

We consider static spherically symmetric geometries which modify the Schwarzschild solution by regularizing the central singularity. A prominent class of solutions propose such generalization by replacing the black hole interior with an asymptotically de Sitter core. We restrict ourselves to solutions with the general form of the metric

\begin{equation}\label{metric_reg}
    d s^2=-\left(1-\frac{2 m(r,\ell)}{r}\right) d t^2+\frac{d r^2}{\left(1-\frac{2 m(r,\ell)}{r}\right)}+r^2 \left(d\theta^2 + \sin^2\theta \phi^2\right),
\end{equation}
where $m(r, \ell)$ is the mass function. The mass function depends on a scale parameter $\ell$ which enables the regularization of the curvature singularity. It encodes the deviation from general relativity and the phenomenological manifestation of quantum gravity effects. For vanishing scale parameter $\ell$ the mass function coincides with the mass of the Schwarzschild black hole. For large radii it behaves asymptotically as the Schwarzschild solution with the same ADM mass up to the leading order terms. Its behavior becomes distinctive at small radii where instead of leading to a curvature singularity it approaches asymptotically a de Sitter core. Consequently, we have the expansion $m(r, \ell)\sim m r^3$ as $r\rightarrow 0$, where $m$ is the AMD mass of the solution. Calculating the mass density we see that it tends to a constant at the coordinate origin. 

Such a regular solution was constructed originally by Bardeen \cite{Bardeen:1968} obtaining the mass function

\begin{equation}
    m(r, \ell)_{Bardeen}=\frac{m r^3}{\left[r^2+\left(2 m \ell^2\right)^{\frac{2}{3}}\right]^{\frac{3}{2}}}.
\end{equation}

It satisfies the Einstein equations coupled to nonlinear electrodynamics (NED) as the scale parameter $\ell$ is connected to the magnetic charge \cite{AyonBeato:1998}. Hayward \cite{Hayward:2006} proposed a related geometry which was further generalized by Frolov and Zelnikov \cite{Frolov:2016}-\cite{Frolov:2017} leading to the mass functions

\begin{eqnarray}
   m(r, \ell)_{Hayward} = \frac{m r^3}{r^3+2 m \ell^2}, \quad\,\,\,\,\,
   m(r, \ell)_{Frolov} = \frac{2 m r^3}{r^3+2 m \ell^2+\ell^3}.
\end{eqnarray}

These metrics can be also associated with solutions to the Einstein-NED equations with an appropriately reconstructed Lagrangian for the electromagnetic field. For our purposes we will consider the geometries as phenomenological models providing regularization of the central singularity regardless of their physical interpretation.

Calculating the components of the Riemann tensor for the described spacetimes we can prove that all the algebraic curvature invariants are finite at small radii. An important distinction is that the mass density and the curvature invariants at the center for the Hayward spacetime are independent of the ADM mass which is not satisfied for the rest. This leads to the existence of a uniform bound on the central algebraic curvature invariants for a fixed regularization length $\ell$ as the mass varies. It is possible to obtain a similar universal bound also for the Frolov spacetime, while for the Bardeen geometry such constraint does not exist.

The curvature singularity of a static spherically symmetric black hole can be regularized by a similar procedure by introducing a central region which tends asymptotically to the Minkowski spacetime at small radii. Such regular spacetimes with asymptotically Minkowski cores were proposed by Simpson and Visser in \cite{Simpson:2019}. They are described by the general form of the metric given by Eq. ($\ref{metric_reg}$) with an exponential mass function

\begin{equation}
   m(r, \ell)_{Simpson} = m e^{-\ell/ r}.
\end{equation}

In contrast to the regular spacetimes with de Sitter core the mass density and the algebraic curvature invariants vanish at the coordinate origin so that the geometry becomes asymptotically locally flat at small radii. The matter density and the curvature peak at a finite distance leaving a hollow core unlike the previously discussed geometries which have a maximum of the mass distribution at the center. Similar to the Bardeen spacteime we cannot impose an uniform curvature bound for the Simpson-Visser geometry since the curvature invariants can grow unlimitedly for a fixed regularization parameter as the ADM mass increases.

It is convenient to introduce a dimensionless regularization parameter by making the rescaling $\ell \rightarrow m\ell$ in the metric of the described spacetimes. Depending on the value of the dimensionless parameter $\ell$ they represent compact objects with different physical nature. For low values of $\ell$ we obtain generalizations of the Schwarzschild black hole with regularized central singularity.  Two Killing horizons are present for scale parameter below some critical value $\ell_h$. They merge for $\ell=\ell_h$ leading to extremal black holes and disintegrate for higher values. At this range the structure of the circular geodesics resembles that of the Schwarzschild black hole containing a single unstable photon sphere, a marginally bound, and a marginally stable timelike orbit.

The geometries with scale parameter $\ell>\ell_h$ describe regular particle-like solutions. They are ultracompact until a second critical value of the regularization parameter $\ell=\ell_{ph}$ possessing a stable (inner) and an unstable (outer) photon spheres. For values higher than  $\ell=\ell_{ph}$ the geometries contain no light rings. The particle-like spacetimes further exhibit a more complicated structure of the timelike circular orbits. Two families of marginally bound circular orbits (MBCO) exist for values of the regularization parameter $\ell>\ell_{h}$ until a critical value $\ell=\ell_{mb}$ where the two families merge. Geometries with $\ell>\ell_{mb}$ possess no MBCO and all the timelike circular geodesics are bound. Similarly, for values of the regularization parameter $\ell>\ell_{ph}$ two families of marginally stable circular orbits (MSCO) exist merging at the critical value $\ell=\ell_{ms}$. Geometries with $\ell>\ell_{ms}$ contains only stable timelike circular orbits. An important characteristic of the particle-like solutions is that there exists an innermost circular orbit (ICO). It corresponds to a timelike circular orbit with a vanishing angular momentum as measured by an asymptotic observer and below this limit timelike circular orbits do not exist.

\begin{figure}[t!]
    \centering
    \includegraphics[width=10.0cm]{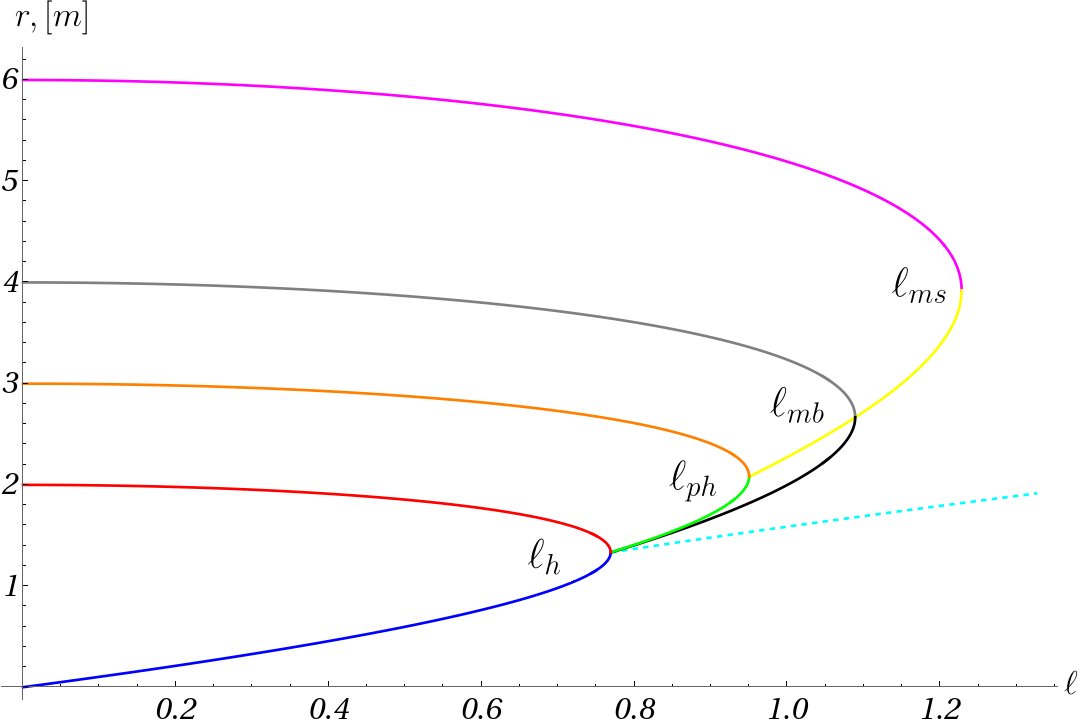}
    \caption{{Characteristic surfaces and circular orbits for the Hayward spacetime. The two Killing horizons (blue and red lines) exist for $\ell< \ell_h$ and merge at $\ell=\ell_h=0.77$. The stable (green) and unstable (orange) photon spheres merge at $\ell= \ell_{ph}=0.95$. The two families of MBCOs and MSCOs are represented in black/grey, and yellow/purple, respectively, merging at  $\ell = \ell_{mb}=1.09$  and $\ell=\ell_{ms}=1.23$. The dashed cyan line represents the location of the ICOs.  For the ultracompact particle-like geometries the timelike circular orbits are stable between the ICO and the inner  photon sphere and unstable between the outer photon sphere and the MSCO. In the region between the two families of photon spheres timelike circular orbits do not exist.}
    \label{Hayward_str}}
\end{figure}

The described structure is common for all the spacetimes which we consider. Therefore, we illustrate in Fig. $\ref{Hayward_str}$ the behavior of the characteristic surfaces and circular orbits only for the Hayward geometry as a representative example. Differences arise in the values of the critical scale parameters for each spacetime which are listed in Table $\ref{crit_str}$.

\begin{table}[]
    \centering
      \begin{tabular}{|c|c|c|c|c|}
       \hline
       \thead{Critical scale parameter }&{{Hayward}}&{{Bardeen}}&{{Frolov}}&{{Simpson}}
          \\  \hline
          
         \thead{ $\ell_h$ (horizons merge)}   &  \thead{$0.77$}&  \thead{$0.48$}&  \thead{$0.67$}&  \thead{$0.74$}
          \\  \hline

                   \thead{ $ \ell_{ph}$ (photon spheres merge)}   &  \thead{$0.95$}&  \thead{$0.56$}&  \thead{$0.80$}&  \thead{$0.80$}
          \\  \hline

                   \thead{ $\ell_{mb}$ (marginally bound orbits merge)}   &  \thead{$1.09$}&  \thead{$0.61$}&  \thead{$0.89$}&  \thead{$0.82$}
          \\  \hline

                   \thead{ $\ell_{ms}$ (marginally stable orbits merge)}   &  \thead{$1.23$}&  \thead{$0.66$}&  \thead{$1.00$}&  \thead{$0.85$}
          \\  \hline

       \end{tabular}
      \caption{Critical values of the scale parameter $\ell$ for the different regular black holes and particle-like geometries.
      \label{crit_str}}      
    \end{table}

\begin{figure}[t!]
    \centering
    \includegraphics[width=10.0cm]{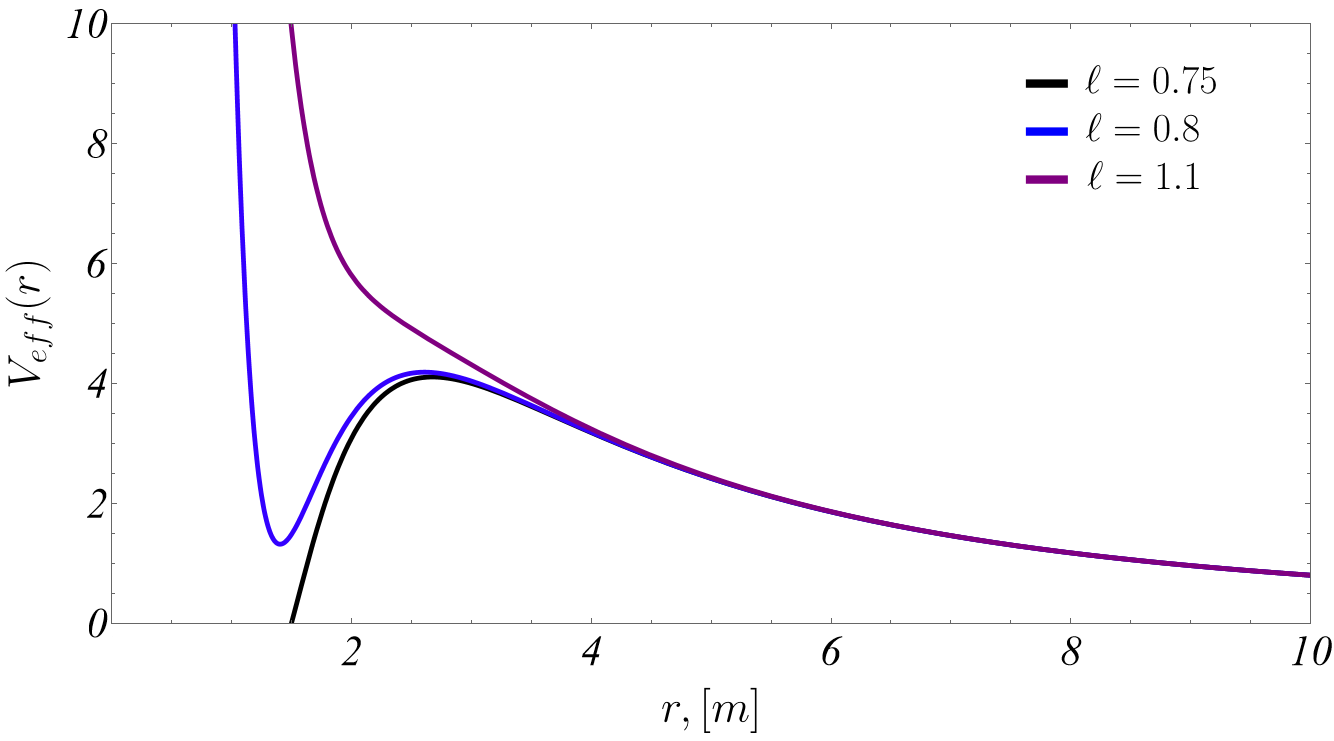}
    \caption{{Behavior of the effective potential for the null geodesics for the Hayward spacetime. We illustrate the different physical regimes by showing regular black holes with scale parameter $\ell = 0.75$ in black, ultracompact particle-like solutions with scale parameter $\ell = 0.8$ in blue, and particle-like solutions possessing no light rings with $\ell = 1.1$ in purple. }
    \label{fig:Hayward_potential}}
\end{figure}

In this work we are interested in the propagation of the null geodesics in the regular geometries. Therefore, it is important to consider the properties of the effective potential which governs the possible qualitative behavior of the photon trajectories. Static spherically symmetric spacetimes with metric given in Eq. ($\ref{metric_reg}$) are characterized by an effective potential for the null geodesics

\begin{equation}
V_{eff} =\frac{ L^2}{r^2}\left(1 - \frac{2m(r,\ell)}{r}\right).
\end{equation}
For the regular black holes which we consider the effective potential behaves in a similar way as for the Schwarzschild black hole possessing a single maximum corresponding to the unstable photon sphere. For the ultracompact particle-like solutions it possesses a inner minimum and an outer maximum corresponding to the stable and unstable photon spheres, respectively, while for the particle-like solutions with no light rings it is monotonous. A characteristic feature for the particle-like solutions is that the effective potential diverges for small radial distances creating a potential barrier which reflects the photon trajectories. Such behavior is expected since the quantum gravity effects needed for the regularization of the singularity are supposed to create a repulsive gravitational field. This type of effective potential also appears for repulsive naked singularities leading to characteristic signatures in the gravitational lensing \cite{Delijski:2023}, \cite{Nedkova:2020}-\cite{Nedkova:2021}. In Fig. $\ref{fig:Hayward_potential}$ we illustrate the effective potential for the three different physical regimes. Its qualitative behavior is common for all the regular spacetimes which we consider. Therefore, we showcase only the Hayward spacetime as a representative example.

\section{Polarized image of an equatorial emitting ring}

In this section, we construct the polarized image of
synchrotron emission originating from a geometrically thin equatorial
emitting ring in the regular spacetimes considered above. The observables of
interest are the polarized intensity and the orientation of the linear
polarization on the observer's sky. The formalism follows the approach
developed in Refs.~\cite{Delijski:2023,Nedkova:2023}, and only the ingredients
required for the present analysis are summarized here.

We introduce an orthonormal tetrad adapted to the static,
spherically symmetric geometry,
\begin{equation}\label{eq:tetrad}
e_{(t)}=(-g_{tt})^{-1/2}\partial_t,\qquad
e_{(r)}=g_{rr}^{-1/2}\partial_r,\qquad
e_{(\theta)}=\frac{1}{r}\partial_\theta,\qquad
e_{(\phi)}=\frac{1}{r\sin\theta}\partial_\phi .
\end{equation}
The radial basis vector is chosen to point outward, toward increasing $r$. The components of the photon four-momentum measured in this
local orthonormal frame are obtained by projection onto the tetrad. For the
metric considered here, they are
\begin{equation}\label{tetrad_momenta}
p^{(t)}=\frac{E}{\sqrt{-g_{tt}}},
\qquad
p^{(r)}=\frac{p_r}{\sqrt{g_{rr}}},
\qquad
p^{(\theta)}=\frac{p_\theta}{r},
\qquad
p^{(\phi)}=\frac{L}{r\sin\theta},
\end{equation}
where $E=-p_t$ and $L=p_\phi$ denote the conserved photon
energy and azimuthal angular momentum, respectively.

The local motion of the emitting plasma is parametrized by
\begin{equation}\label{eq:betaboost}
\vec{\beta}
=
\beta\left(
\cos\chi\,e_{(r)}
+\sin\chi\,e_{(\phi)}
\right),
\end{equation}
where $\beta$ is the magnitude of the plasma three-velocity
in units of the speed of light and $\chi$ determines its direction in the
equatorial plane. The velocity field is treated phenomenologically and is not assumed, in general, to coincide with the four-velocity of matter on a
circular geodesic.

The tetrad frame \eqref{eq:tetrad} is related to the rest
frame of the emitting fluid by a local Lorentz boost with velocity
$\vec{\beta}$. In the basis $(t,r,\phi,\theta)$, the corresponding
transformation is
\begin{equation}\label{eq:lorentz_boost}
\hat e_{(a)}^{\,\mu}
=
\Lambda_{(a)}^{\;\;\;\,(b)}e_{(b)}^{\,\mu},
\end{equation}
\vspace{1.0mm}
\begin{equation}\label{eq:lorentz_matrix}
\begin{gathered}
\Lambda
=
\begin{pmatrix}
\gamma
&
-\beta\gamma\cos\chi
&
-\beta\gamma\sin\chi
&
0
\\[1mm]
-\beta\gamma\cos\chi
&
1+(\gamma-1)\cos^2\chi
&
(\gamma-1)\sin\chi\cos\chi
&
0
\\[1mm]
-\beta\gamma\sin\chi
&
(\gamma-1)\sin\chi\cos\chi
&
1+(\gamma-1)\sin^2\chi
&
0
\\[1mm]
0&0&0&1
\end{pmatrix},
\\[2mm]
\gamma=(1-\beta^2)^{-1/2}.
\end{gathered}
\end{equation}
Since the boost is confined to the equatorial
$(r,\phi)$ plane, the $\theta$ direction is unchanged. In the following,
hatted components refer to quantities measured in the co-moving fluid frame.

The magnetic field and photon three-momentum in this frame
are written as
$\vec{B}=(\hat{B}^{r},\hat{B}^{\phi},\hat{B}^{\theta})$ and
$\vec{p}=(\hat{p}^{r},\hat{p}^{\phi},\hat{p}^{\theta})$, respectively.

For synchrotron emission, the local polarization direction
is orthogonal to both the photon propagation direction and the magnetic
field, and is therefore represented by
\begin{equation}\label{eq:polcross}
\vec{f}
=
\frac{\vec{p}\times\vec{B}}{|\vec{p}|}.
\end{equation}
The polarization vector has vanishing temporal component in
the local fluid frame and satisfies
\begin{equation}
\hat{f}^{a}\hat{f}_{a}
=
\sin^2\zeta\,|\vec{B}|^2,
\qquad
\sin\zeta
=
\frac{|\vec{p}\times\vec{B}|}
{|\vec{p}||\vec{B}|},
\end{equation}
where $\zeta$ is the pitch angle between the photon momentum
and the magnetic field.

To obtain the polarization measured by a distant observer,
the polarization four-vector must be parallel transported along the
corresponding null geodesic. Together with the null-geodesic equation, it
satisfies
\begin{equation}
p^\mu\nabla_\mu p^\nu=0,
\qquad
p^\mu\nabla_\mu f^\nu=0,
\qquad
p_\mu f^\mu=0.
\end{equation}
For the static and spherically symmetric metrics considered
here, the polarization transport can be expressed algebraically by exploiting the hidden symmetries associated with a rank-two Killing--Yano tensor. For the metric class with $g_{tt}g_{rr}=-1$, its only independent non-vanishing component can be written as
\begin{equation}\label{eq:KY_tensor}
Y_{\theta\phi}
=
-Y_{\phi\theta}
=
r^3\sin\theta .
\end{equation}
Its Hodge dual,
$\widetilde{Y}_{\mu\nu}
=\frac{1}{2}\epsilon_{\mu\nu\alpha\beta}Y^{\alpha\beta}$,
is a conformal Killing--Yano tensor with
\begin{equation}\label{eq:CKY_tensor}
\widetilde{Y}_{tr}
=
-\widetilde{Y}_{rt}
=
r .
\end{equation}
These hidden symmetries give rise to two conserved
quantities along the null geodesic,
\begin{equation}\label{kappa_def}
\kappa_1
=
\widetilde{Y}_{\mu\nu}p^\mu f^\nu,
\qquad
\kappa_2
=
Y_{\mu\nu}p^\mu f^\nu ,
\end{equation}

which, for the present geometries, reduce to
\begin{equation}\label{kappa_12}
\begin{aligned}
\kappa_1
&=
r\left(p^t f^r-p^r f^t\right),
\\[1mm]
\kappa_2
&=
r^3\sin\theta
\left(p^\theta f^\phi-p^\phi f^\theta\right).
\end{aligned}
\end{equation}
The conservation of $\kappa_1$ and $\kappa_2$ provides an
algebraic link between the polarization at the emission point and that
measured by the distant observer. In the numerical implementation, these
constants are evaluated from the photon momentum and polarization vector at
the emission point, using the local Lorentz transformation
\eqref{eq:lorentz_boost}--\eqref{eq:lorentz_matrix}, and remain fixed along
the corresponding null geodesic. This avoids the explicit numerical
integration of the polarization parallel-transport equations.

At the observer position $(r_0,\theta_0)$, the direction
of the incoming photon is parametrized in the local orthonormal frame
\eqref{eq:tetrad} by the observation angles $\tilde{\alpha}$ and $\tilde{\beta}$, defined
through
\begin{equation}\label{obs_angles}
\sin\tilde{\alpha}
=
\frac{p^{(\theta)}}{p^{(t)}},
\qquad
\tan\tilde{\beta}
=
\frac{p^{(\phi)}}{p^{(r)}}.
\end{equation}
The corresponding Cartesian coordinates on the observer's
image plane are
\begin{equation}\label{obs_coord}
\begin{aligned}
x
&=
-r_0\tan\tilde{\beta}
=
-r_0\,\frac{p^{(\phi)}}{p^{(r)}}
\simeq
-\frac{p_\phi}{E\sin\theta_0}
=
-\frac{L}{E\sin\theta_0},
\\[2mm]
y
&=
r_0\sin\tilde{\alpha}
=
r_0\,\frac{p^{(\theta)}}{p^{(t)}}
\simeq
\frac{p_\theta}{E}.
\end{aligned}
\end{equation}
In obtaining the asymptotic expressions, we used
$p^{(t)}\to E$ from asymptotic flatness and $p^{(r)}\to E$ from the null condition for photons propagating outward toward the distant observer, consistently with the radial tetrad orientation adopted in Eq. \eqref{eq:tetrad}. Thus, $\tilde{\alpha}$ and $\tilde{\beta}$ specify the local photon direction, while $(x,y)$ determine the corresponding position on the observer's image plane.

The conserved quantities $\kappa_1$ and $\kappa_2$ then
determine the polarization projected onto the observer's image plane.
Before normalization, the corresponding components are

\begin{equation}\label{obs_polarization_unnorm}
\begin{aligned}
f^x
&=
\frac{x\kappa_1+y\kappa_2}{x^2+y^2},
\\[1mm]
f^y
&=
\frac{y\kappa_1-x\kappa_2}{x^2+y^2}.
\end{aligned}
\end{equation}
For convenience, we normalize the projected polarization
vector to unit magnitude, obtaining
\begin{equation}\label{obs_polarization}
\begin{aligned}
f^x_{\rm obs}
&=
\frac{x\kappa_1+y\kappa_2}
{\sqrt{(\kappa_1^2+\kappa_2^2)(x^2+y^2)}},
\\[2mm]
f^y_{\rm obs}
&=
\frac{y\kappa_1-x\kappa_2}
{\sqrt{(\kappa_1^2+\kappa_2^2)(x^2+y^2)}}.
\end{aligned}
\end{equation}
Eqs.~\eqref{kappa_12} and \eqref{obs_polarization} therefore provide the direct connection between the local polarization generated at the emitting ring and its orientation on the observer's sky.

The synchrotron emission intensity depends on the pitch
angle $\zeta$, the photon frequency, and the effective photon path length
through the emitting medium \cite{Narayan:2021}, given by
\begin{equation}\label{eq:path_length}
l_p
=
\frac{\hat p^{(t)}}{\hat p^{(\theta)}}\,H,
\end{equation}
where $H$ is the vertical thickness of the geometrically
thin emitting region, while $\hat p^{(t)}$ and $\hat p^{(\theta)}$ denote
the photon four-momentum components measured in the rest frame of the
emitting fluid. In this prescription, $l_p$ provides an effective estimate
of the photon path length through the emitting medium within the
geometrically thin, optically thin approximation
\cite{Narayan:2021}.

This prescription forms part of a simplified geometrical
model of polarized synchrotron emission, whose purpose is to isolate the
effects of photon propagation and polarization transport in the underlying
spacetime rather than to provide a self-consistent radiative-transfer model
of the accretion flow \cite{Narayan:2021,Gelles:2021}.

For the distant observer considered above, the photon
energy is $E$, while in the co-moving frame of the emitting plasma it is
$\hat p^{(t)}$. The corresponding frequency-shift factor is therefore
\begin{equation}\label{eq:doppler_factor}
\delta
=
\frac{\nu_{\rm obs}}{\nu_{\rm em}}
=
\frac{E}{\hat p^{(t)}} .
\end{equation}
The polarized intensity then acquires the standard Doppler
weighting $\delta^{3+\alpha_\nu}$, where $\alpha_\nu$ is the spectral index. The resulting expression is \cite{Narayan:2021}
\begin{equation}\label{eq:polarized_intensity}
|I|
=
\delta^{3+\alpha_\nu}\,
l_p\,
(\sin\zeta)^{1+\alpha_\nu}.
\end{equation}

The spectral index $\alpha_\nu=1$ is adopted for
consistency with the polarized synchrotron model used in applications to
M87* \cite{Narayan:2021,Gelles:2021}.

The observed electric-field components associated with the
linear polarization are
\begin{equation}\label{pol_EVPA}
\begin{aligned}
E^x_{\rm obs}
&=
\delta^2\,l_p^{1/2}\,\sin\zeta\,f^x_{\rm obs},
\\[2mm]
E^y_{\rm obs}
&=
\delta^2\,l_p^{1/2}\,\sin\zeta\,f^y_{\rm obs},
\\[2mm]
|I|
&=
(E^x_{\rm obs})^2+(E^y_{\rm obs})^2.
\end{aligned}
\end{equation}

The electric-vector position angle (EVPA) specifies the
orientation of the linear polarization on the observer's sky and is measured
counterclockwise from the positive $x$-axis,
\begin{equation}\label{eq:EVPA}
{\rm EVPA}
=
\arctan\left(-\frac{f^x_{\rm obs}}{f^y_{\rm obs}}\right).
\end{equation}
For linearly polarized radiation, angles differing by
$\pi$ represent the same physical polarization state, and the EVPA is
therefore defined modulo $\pi$. The expression above selects a particular
branch for its numerical representation, which is used consistently
throughout the analysis. Apparent jumps associated solely with the angular
branch convention should consequently be distinguished from genuine changes
in the polarization orientation.

\section{Images of circular orbits}

In order to obtain the observable linear polarization of an emitting thin fluid ring we need to construct the image of the equatorial ring on the observer's sky and calculate the pattern of the polarization vector around it. The first step of the procedure reduces to studying the images of the circular orbits in the particular spacetime. For static spherically symmetric geometries this analysis can be done semi-analytically giving us intuition about the image morphology and possible observational signatures. The method combines the integral form of the null-geodesic equations with a numerical determination of the photon trajectories reaching a distant observer. A detailed derivation of the procedure is presented in Refs.~\cite{Nedkova:2019}-\cite{Nedkova:2020}. Here we retain only the elements required for the present analysis.

We consider a general static and spherically symmetric spacetime described by
\begin{equation}
ds^2=-A_1(r)\,dt^2+A_2(r)\,dr^2
+r^2\left(d\theta^2+\sin^2\theta\,d\phi^2\right),
\end{equation}
where, for the geometries studied in this work, $A_1(r)=1/A_2(r)$. The observable image of a circular equatorial emitter is determined by the null-geodesic structure of the underlying spacetime. In particular, strong gravitational deflection and the possible existence of radial turning points determine the number, location, and morphology of the
image components reaching the observer.

Exploiting spherical symmetry, each photon trajectory can be described in its orbital plane. The accumulated azimuthal angle along a null geodesic with impact parameter $b=J/E$ is given by
\begin{equation}\label{phi:1}
\Delta\phi(b) = \int
\frac{dr}
{r^2\sqrt{\frac{1}{b^2}
-\frac{A_1}{r^2}}}.
\end{equation}
Here $J$ denotes the conserved magnitude of the photon angular momentum, $J^2=p_\theta^2+L^2/\sin^2\theta$. In the asymptotically flat region, the impact parameter satisfies $b=\sqrt{x^2+y^2}$, consistently with Eq. (\ref{obs_coord}). We consider photons emitted from a circular equatorial orbit at $r=r_{\rm source}$ and received by an observer located at $r=r_{\rm obs}$.

To relate a photon trajectory to its apparent position on the observer's sky, we use the angles $\tilde{\alpha}$ and $\tilde{\beta}$ introduced in Eq. \eqref{obs_angles}. Denoting the photon energy measured by the static observer by $\mathcal{E}_{\rm obs}=p^{(t)}=E/\sqrt{-g_{tt}(r_{\rm obs})}$, we write the covariant spatial momentum components at the observer as
\begin{eqnarray}\label{p_alpha}
&&p_\theta=\mathcal{E}_{\rm obs}\sqrt{g_{\theta\theta}}\,\sin\tilde{\alpha},
\qquad
p_\phi=\mathcal{E}_{\rm obs}\sqrt{g_{\phi\phi}}\,
\sin\tilde{\beta}\,\cos\tilde{\alpha},
\nonumber\\[2mm]
&&p_r=\mathcal{E}_{\rm obs}\sqrt{g_{rr}}\,
\cos\tilde{\beta}\,\cos\tilde{\alpha},
\end{eqnarray}
where all metric components are evaluated at the observer's position. For a fixed initial direction, changing $\mathcal{E}_{\rm obs}$ rescales the photon frequency without altering the null trajectory. We therefore set $\mathcal{E}_{\rm obs}=1$ when constructing the images. Together with Eq. \eqref{obs_coord}, these relations connect the local photon direction to the corresponding position on the image plane.

It is convenient to introduce two further celestial angles, $\eta$ and $\xi$, in the observer's local frame \eqref{eq:tetrad}. The angle $\eta$ specifies the orientation of the photon orbital plane with respect to the azimuthal direction, whereas $\xi$ measures the angular separation between the photon propagation direction and the radial direction toward the compact object. In the asymptotic region, the corresponding celestial coordinates $\tilde{\alpha}$ and $\tilde{\beta}$ are related to these angles by
\begin{equation}\label{phi:celecele}
\tilde{\alpha}=\xi\cos\eta,
\qquad
\tilde{\beta}=\xi\sin\eta.
\end{equation}

Spherical symmetry further allows the azimuthal coordinate of the emission point to be related to the observer inclination $i$ and the celestial angle $\eta$. With the convention adopted here, this relation takes the form
\begin{equation}\label{phi:2}
\phi
=
-\arccos
\left(
\frac{\sin\eta\,\tan i}
{\sqrt{1+\sin^2\eta\,\tan^2 i}}
\right).
\end{equation}

Combining Eqs.~\eqref{phi:1} and \eqref{phi:2} yields a relation between the impact parameter $b$ and the celestial direction $\eta$ for photons emitted at $r_{\rm source}$ and received by an observer at $(r_{\rm obs},i)$. For trajectories that connect the source and observer without a radial turning point, one obtains
\begin{equation}\label{orbit:1}
\int_{r_{\rm source}}^{r_{\rm obs}}
\frac{dr}
{r^2\sqrt{\frac{1}{b^2}
-\frac{A_1}{r^2}}}
=
-\arccos
\left(
\frac{\sin\eta\,\tan i}
{\sqrt{1+\sin^2\eta\,\tan^2 i}}
\right).
\end{equation}

More generally, photon trajectories may accumulate additional angular winding around the compact object before reaching the observer, thereby generating higher-order images. These trajectories are described by
\begin{equation}\label{orbit:2}
\int_{r_{\rm source}}^{r_{\rm obs}}
\frac{dr}
{r^2\sqrt{\frac{1}{b^2}
-\frac{A_1}{r^2}}}
=
k\pi
-
\arccos
\left(
\frac{\sin\eta\,\tan i}
{\sqrt{1+\sin^2\eta\,\tan^2 i}}
\right),
\end{equation}
where the integer $k$ labels the image order according to the convention adopted in the present construction.

Photon trajectories can be separated into two classes according to whether or not they possess a radial turning point. For a trajectory with a turning point at $r=r_0$, the photon initially propagates toward smaller radii, reaches $r_0$, and subsequently returns to larger radii before arriving at the observer. The corresponding radial integral then separates into two branches,
\begin{equation}\label{orbit:11}
\int_{r_0}^{r_{\rm source}}
\frac{dr}
{r^2\sqrt{\frac{1}{b^2}
-\frac{A_1}{r^2}}}
+
\int_{r_0}^{r_{\rm obs}}
\frac{dr}
{r^2\sqrt{\frac{1}{b^2}
-\frac{A_1}{r^2}}}
=
k\pi
-
\arccos
\left(
\frac{\sin\eta\,\tan i}
{\sqrt{1+\sin^2\eta\,\tan^2 i}}
\right).
\end{equation}

At the radial turning point, the radial momentum vanishes, which relates the impact parameter to the turning-point radius according to
\begin{equation}\label{eq:b9}
b=\frac{r_0}{\sqrt{A_1(r_0)}}.
\end{equation}
In the limiting case in which the turning point approaches the emitting radius, $r_0\rightarrow r_{\rm source}$, the corresponding limiting value of the impact parameter becomes
\begin{equation}\label{eq:b10}
b_{\rm lim}
=
\frac{r_{\rm source}}
{\sqrt{A_1(r_{\rm source})}}.
\end{equation}

\begin{figure}[t!]
    		\setlength{\tabcolsep}{0pt}{\small\tt
		\begin{tabular}{cc}
           \includegraphics[width=0.45\textwidth]{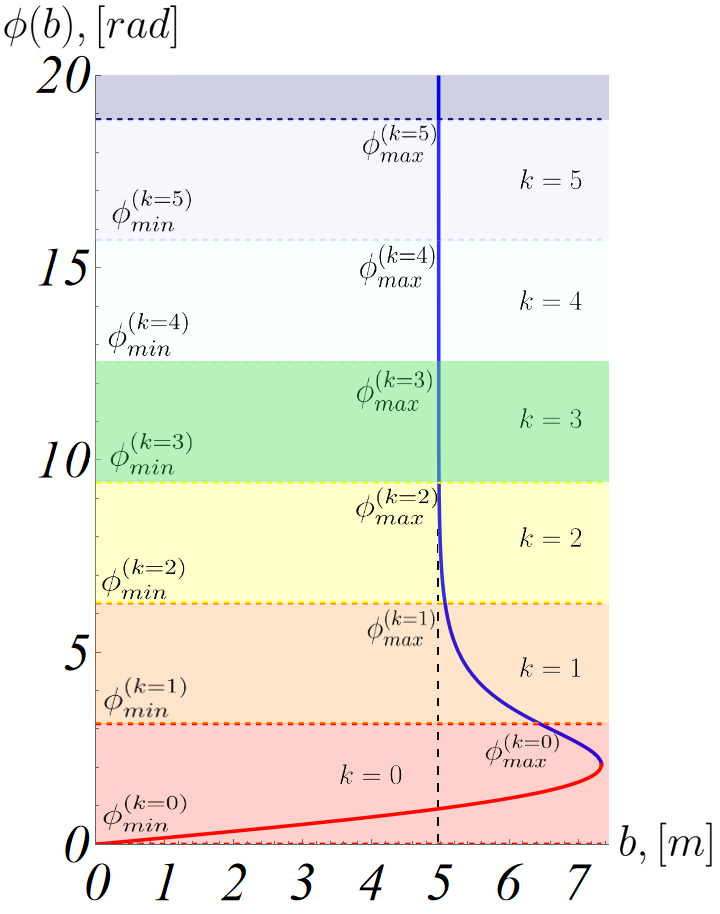}
		   \includegraphics[width=0.5\textwidth]{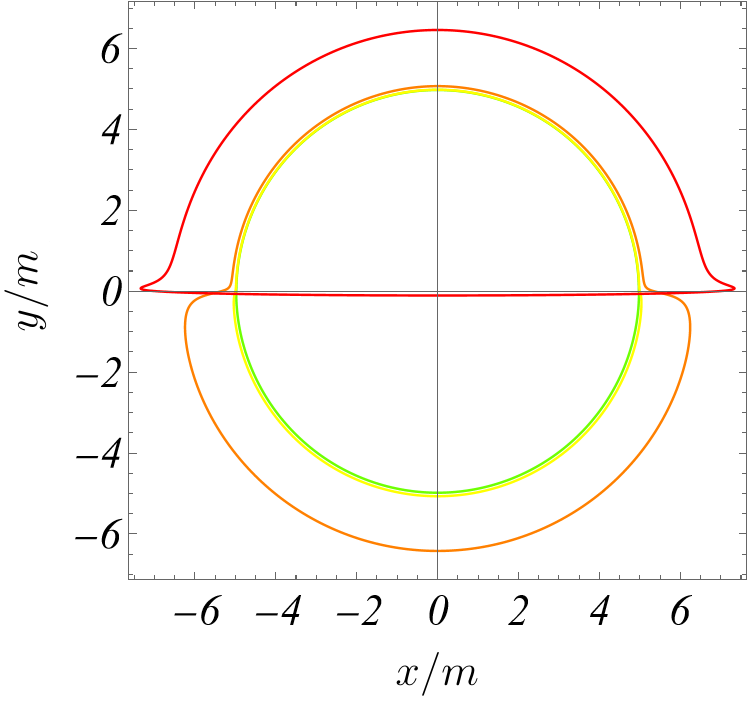} \\[1mm]
           \hspace{0.7cm} $a)$ \hspace{7.5cm} $b)$
        \end{tabular}}
\caption{\label{fig:Image_reg}
Deflection curve and corresponding images of a circular
equatorial source for the Hayward regular black hole. The regularization parameter is  $\ell=0.7$. Panel (a) shows the relevant deflection
curve branches and observational intervals, while panel (b) displays the
corresponding images on the observer's sky. We depict with the same color the observational interval $\Delta\phi^{(k)}$ in panel a),  and the corresponding image of order $k$ in panel b). The vertical dashed line denotes the location of the photon sphere. Images of order $k>2$ approach closely the shadow boundary and become hard to distinguish. The source is located at $r_{\rm source}=6m$, and the observer is placed at $r_{\rm obs}=10^4m$
with inclination $i=89^\circ$.}
\end{figure}

\begin{figure}[t!]
    		\setlength{\tabcolsep}{0pt}{\small\tt
		\begin{tabular}{cc}
         \includegraphics[width=0.45\textwidth]{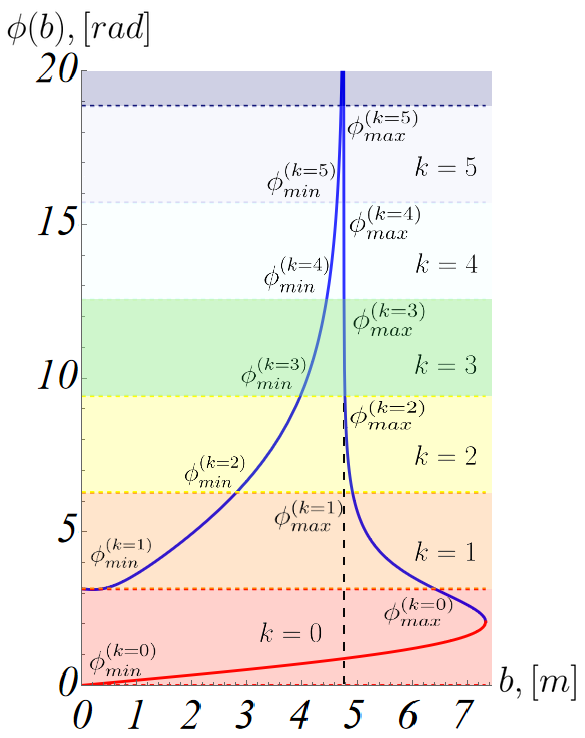}
        \includegraphics[width=0.45\textwidth]{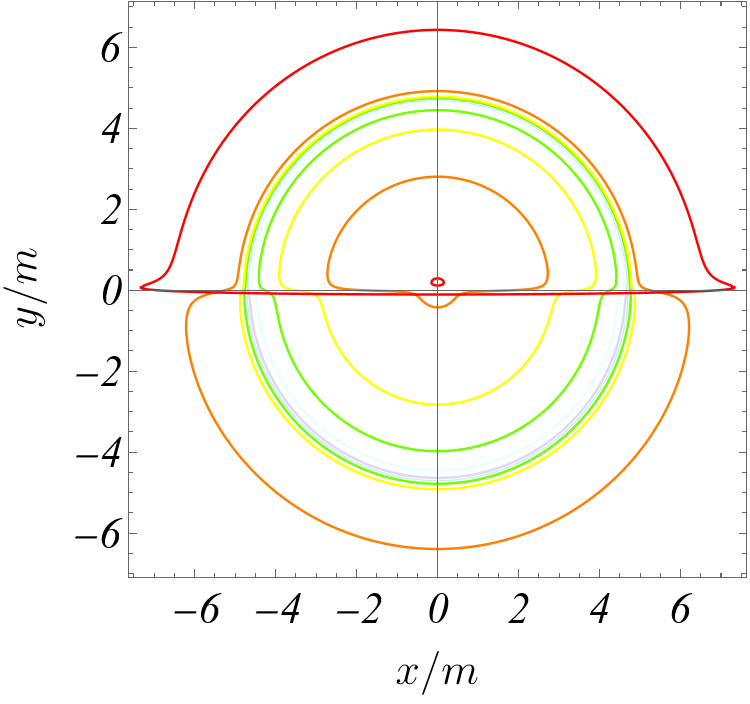} \\[1mm]
          \hspace{7cm} $a)$ \hspace{7cm} \\
        \includegraphics[width=0.45\textwidth]{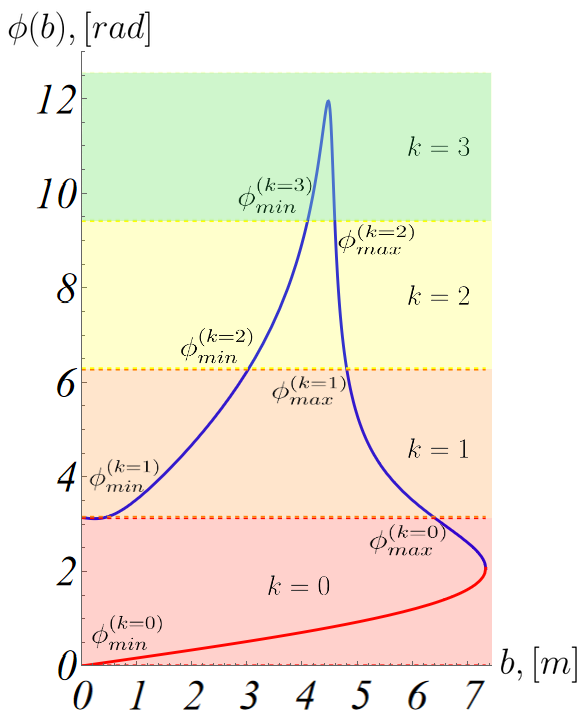}
		   \includegraphics[width=0.48\textwidth]{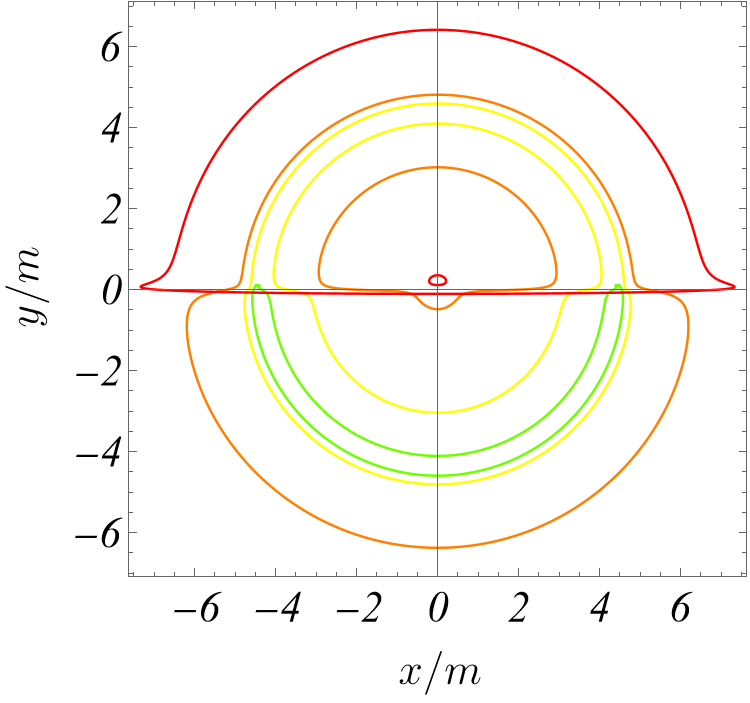} \\[1mm]
         \hspace{7cm} $b)$ \hspace{7cm}
        \end{tabular}}
\caption{\label{fig:Image_part}
Deflection curves and corresponding images of a circular
equatorial source for Hayward particle-like geometries. Panel (a) displays ultracompact solutions with parameter  $\ell=0.9$ , while panel (b) showcases  solutions with no light rings and  $\ell=1$. We follow the same conventions for color representation and source-observer location as in Fig. \ref{fig:Image_reg}.}
\end{figure}

For each image order $k$ and inclination angle $i$, Eq.~\eqref{orbit:2} defines an allowed observational interval as the celestial angle $\eta$ varies over the observer's sky. For a fixed inclination angle we denote its extent by
$\Delta\phi^{(k)}=\phi_{\max}^{k}-\phi_{\min}^{k}$. In order to perform a qualitative analysis of the observable images it is convenient to represent graphically the deflection curve $\phi(b)$ and its superposition with the observational intervals $\Delta\phi^k$ for each image order (see Figs. \ref{fig:Image_reg}-\ref{fig:Image_part}). In this graphical construction the intersections of a given observational interval with the relevant branches of the deflection curve identify photon trajectories contributing to the observable image. The blue curves correspond to trajectories possessing a radial turning point, whereas the red curves represent trajectories without one.

The graphical analysis allows us to predict the morphology of the observable images. Depending on the spacetime parameters and on the source--observer configuration, a given observational interval may intersect more than one disconnected portions of the deflection curve. This leads to the creation of an image of the corresponding order which
contains two spatially disconnected components. On the other hand
if a particular branch of the deflection curve spans only part of the allowed observational interval, the corresponding image component forms a curve which does not close around the coordinate origin on the observer's sky.

\begin{figure}[t]
    		\setlength{\tabcolsep}{0pt}{\small\tt
		\begin{tabular}{cc}
           \includegraphics[width=0.45\textwidth]{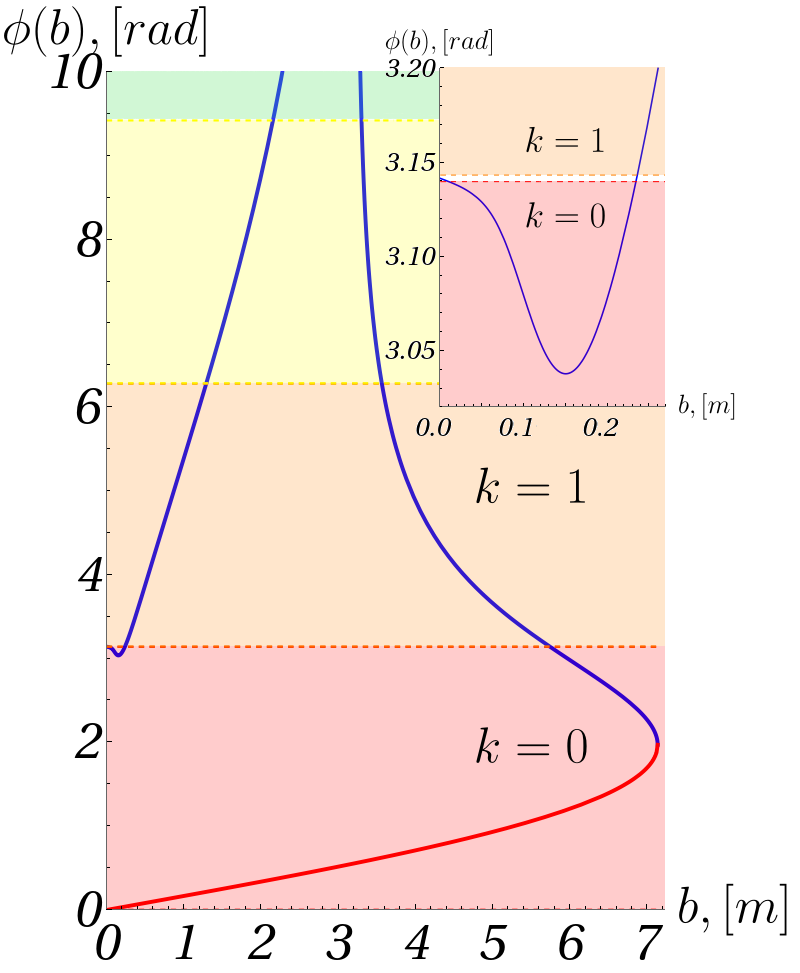}
		   \includegraphics[width=0.45\textwidth]{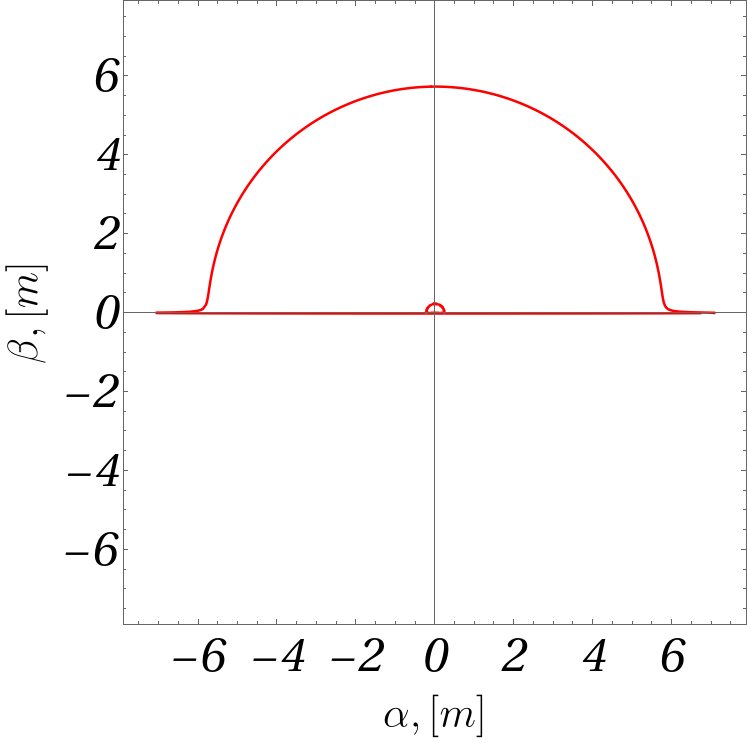} \\[1mm]
           \hspace{0.7cm} $a)$ \hspace{7.5cm} $b)$
        \end{tabular}}
\caption
{Formation of the direct images in for Simpson particle-like ultracompact geometries. The regularization parameter is
$\ell=0.79$. Panel (a) shows the deflection curve for photons emitted from
a circular equatorial source at $r_{\rm source}=6m$ and observed at
$r_{\rm obs}=10^4m$ with inclination $i=89.9^\circ$. The branches
corresponding to trajectories with and without a radial turning point are
shown together with the observational interval for $k=0$. The inner branch of the deflection curve develops a local minimum for small impact parameters which lies in this observational interval, giving rise to a second inner image 
in addition to the conventional outer image. Panel (b) shows the corresponding image components on the observer's sky.}
\label{fig:Image_Simpson}
\end{figure}

In Figs. \ref{fig:Image_reg}-\ref{fig:Image_part} we illustrate the formation of the possible images of the circular orbits in the regular spacetimes which we consider. Our analysis demonstrates that the deflection curves  behave in a qualitatively similar way for all the geometries. Therefore, we showcase the Hayward spacetime as a representative example concentrating on the three regimes in the parameter space with different physical properties. The deflection curve for the regular black hole behaves similarly to the Schwarzschild black hole and leads to the same types of images (see Fig. \ref{fig:Image_reg}). It  intersects each observational interval $\Delta\phi^{(k)}$ for a single connected range of the impact parameter $b$. Therefore, we observe a single image of each order $k$.

\begin{figure}[t]
    		\setlength{\tabcolsep}{0pt}{\small\tt
		\begin{tabular}{cc}
           \includegraphics[width=0.45\textwidth]{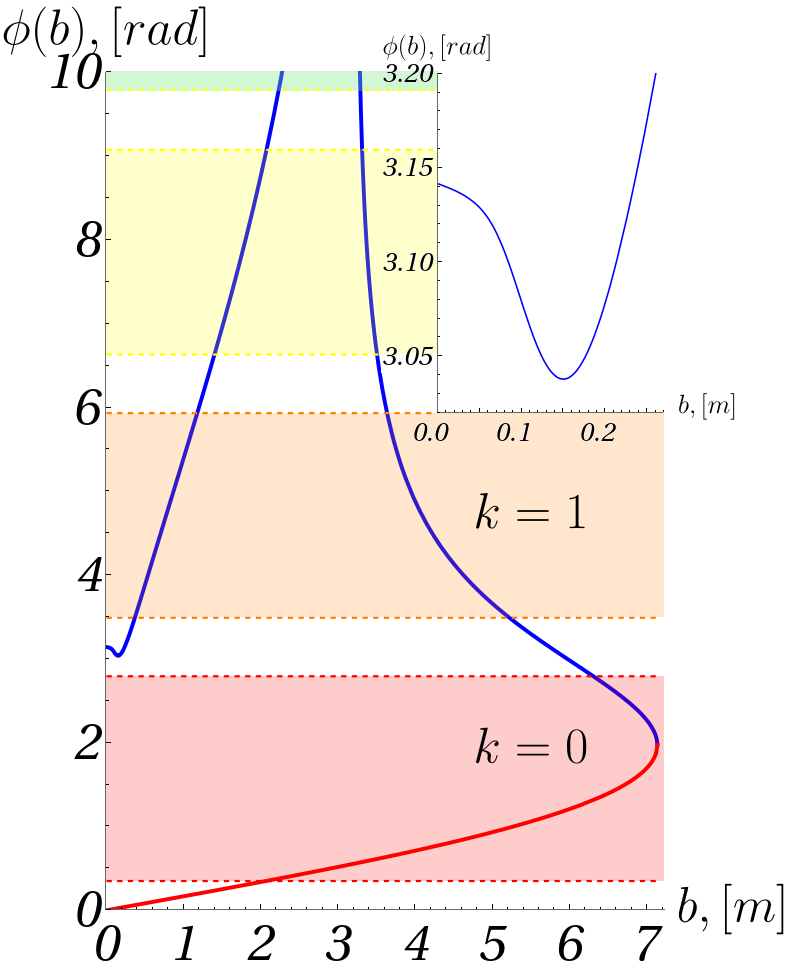}           
		   \includegraphics[width=0.45\textwidth]{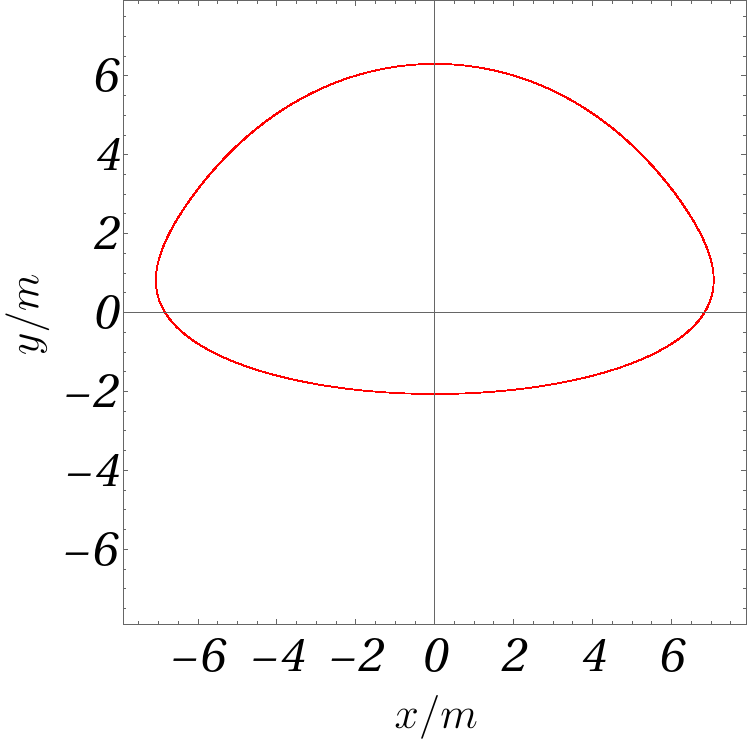} \\[1mm]
           \hspace{0.7cm} $a)$ \hspace{7.5cm} $b)$
        \end{tabular}}
\caption{\label{fig:Image_Simpson1}
Deflection curve and direct image of a circular
equatorial source for the ultracompact particle-like Simpson geometry with regularization parameter $\ell=0.79$.  The source is located at $r_{\rm source}=6m$, and the observer is placed at $r_{\rm obs}=10^4m$
with inclination $i=70^\circ$. The direct inner image does not form due to insufficiently high inclination leading to an observational interval for $k=0$ which does not encompass the local minimum of the deflection curve.}
\end{figure}

More interesting behavior is observed for the particle-like geometries. We observe two branches of the deflection curve corresponding to photon trajectories with a radial turning point. For the ultracompact solutions they approach asymptotically the unstable photon sphere with higher and lower impact parameters, respectively, than the critical impact parameter of the photon sphere (see Fig. \ref{fig:Image_part} a).  For the particle-like geometries  without light rings the two branches merge continuously at the observational interval of the last possible image order, thus forming a global maximum (see Fig. \ref{fig:Image_part} b). In both cases the two deflection branches intersect the observational intervals $\Delta\phi^{(k)}$ for two disconnected ranges of the impact parameter leading to the creation of two spatially disconnected images of order $k$ for all relevant orders. Similar behavior was observed for naked singularities with a repulsive gravitational field at small distances for image orders $k\geq 1$ (see \cite{Nedkova:2020}-\cite{Nedkova:2021}).

In this work we will concentrate on the direct images since they are the most relevant for polarimetric studies. In this respect the particle-like solutions which we consider exhibit a specific lensing feature. The inner branch of the deflection curve develops a local minimum for small impact parameters. For sufficiently large inclinations this minimum resides inside the observational interval of the direct images of order $k=0$. As a result, the direct images contain two spatially separated components: the usual outer image with large impact parameters and an additional compact inner image located in the vicinity of the coordinate origin on the observer's sky . The inner component is produced by photons that propagate toward smaller radii, reach a radial turning point in the vicinity of the regular core, and scatter away to infinity.  A similar additional direct image is absent for the regular black holes and the Schwarzschild black hole since the deflection angle curve in these spacetimes does not contain a second branch located at small impact parameters. More interestingly, such a direct image does not exist for the repulsive naked singularities investigated in \cite{Nedkova:2019}-\cite{Nedkova:2020} which otherwise lead to the formation of a similar structure of double images for higher orders $k\geq1$. Thus, the appearance of an additional inner direct image may turn out to represent a  characteristic phenomenological manifestation of these classes of regular geometries.

In order to visualize more clearly the formation of the inner image, in Fig. $\ref{fig:Image_Simpson}$ we present the deflection curve for the Simpson geometry where the local minimum at small impact parameters is more pronounced than for the Hayward spacetime. Although this example describes an ultracompact particle-like solution, this behavior is preserved also for particle-like solutions with no light rings. The formation of the  inner direct images depends on a combination of parameters including the inclination angle, the radial position of the source and the regularization parameter of the spacetime. Our analysis shows that the effect is characteristic for high inclinations appearing only after a certain critical inclination. While the critical inclination depends on the regularization parameter and the emitting radius, for reasonable values of these parameters we cannot observe the inner direct image for low inclinations such as $i=20^{\circ}$ for example.  The reason is that the width of the observational intervals $\Delta\phi^{(k)}$ depends strongly on the inclination and it increases as the inclination grows. If the inclination is not sufficiently high the observational window for $k=0$ fails to encompass the local minimum of the deflection curve for small impact parameters which is responsible for the formation of the inner direct image. Consequently, such an image is not produced on the observer's sky. In Fig. $\ref{fig:Image_Simpson1}$ we illustrate such an example when the inner direct image is not observed due to a insufficiently high inclination angle. In Section 6 we perform a detailed analysis of the parameter space where direct inner images exist and study the properties of the polarized radiation observed at their location.

\section{Linear polarization for regular black holes and particle-like solutions}

In this section, we investigate the linearly polarized synchrotron emission from geometrically thin equatorial emitting regions in the regular spacetimes introduced in Section 2. We concentrate on low inclinations which are compatible with the physical settings of the EHT galactic targets M87* and Sgr A* restricting the analysis to direct images, i.e. to photon trajectories that reach the observer without additional winding around the compact object. Within the parametrization adopted in Section 4, these trajectories correspond to the azimuthal interval $\phi\in[0,\pi)$. In this regime the circular orbits in the regular spacetimes possess a single image which is qualitatively similar to the black holes in general relativity. This motivates performing a quantitative comparison of the polarization properties with the Schwarzschild black hole and assessing the possibility of distinguishing regular spacetimes by means of the EHT observational data.

In the following analysis we compare the polarized intensity and the EVPA with the corresponding Schwarzschild predictions using the formalism developed in Sections 3 and 4. In order to perform this comparison, we first determine the apparent position on the observer's sky of the image of a circular emitter located at $r_{\rm source}=6m$ in the Schwarzschild spacetime. For each regular black hole geometry, the emission radius is then adjusted so that the corresponding image is formed at the same reference position on the observer's sky.

\begin{table}[t]
    \centering
      \begin{tabular}{|c|c|c|c|c|}
       \hline
       \thead{ $i = 20^{\circ}$ }& \multicolumn{2}{c|}{{$B = [0.87, 0.5, 0]$, $\chi=-150\degree$}} & \multicolumn{2}{c|}{{$B = [0.71, 0.71, 0]$, $\chi=-135\degree$}} 
          \\  \hline
          
         \thead{ $\ell$}   &  \thead{$\left(\frac{\text{max}\,\Delta \text{I}}{\text{I}_\text{Sch}} \,, \, \phi \, \right)$} & \thead{$\left(\frac{\text{max}\,\Delta \text{EVPA}}{\text{EVPA}_\text{Sch}} \,, \, \phi \, \right)$} &  \thead{$\left(\frac{\text{max}\,\Delta \text{I}}{\text{I}_\text{Sch}} \,, \, \phi \, \right)$} & \thead{$\left(\frac{\text{max}\,\Delta \text{EVPA}}{\text{EVPA}_\text{Sch}} \, , \, \phi \, \right)$} 
          \\  \hline
               \multicolumn{5}{|c|} {Hayward}
          \\  \hline
           \thead{$\vspace{0.1mm}\text{$0.75$}$\vspace{0.1mm}}  &  \thead{(0.0107, 0.35$\pi$)} & \thead{(0.0006, 1.15$\pi$)} &  \thead{(0.0093, 0.29$\pi$)} & \thead{(0.0010, 1.03$\pi$)} 
          \\  \hline

          \thead{\vspace{0.1mm}$\text{$0.85$}$\vspace{0.1mm}} &  \thead{(0.0138, 0.35$\pi$)} & \thead{(0.0008, 1.15$\pi$)} &  \thead{(0.0119, 0.29$\pi$)} & \thead{(0.0013, 1.03$\pi$)} 
          \\  \hline

          \thead{\vspace{0.1mm}$\text{$1.3$}$\vspace{0.1mm}}  &  \thead{(0.0320, 0.35$\pi$)} & \thead{(0.0019, 1.15$\pi$)} &  \thead{(0.0277, 0.29$\pi$)} & \thead{(0.0030, 1.03$\pi$)} 
          \\  \hline
                  \multicolumn{5}{|c|} {Simpson}
          \\  \hline
          
           \thead{$\vspace{0.1mm}\text{$0.70$}$\vspace{0.1mm}}  &  \thead{(0.2246, 0.34$\pi$)} & \thead{(0.0106, 1.19$\pi$)} &  \thead{(0.1948, 0.27$\pi$)} & \thead{(0.0161, 1.07$\pi$)} 
          \\  \hline

          \thead{\vspace{0.1mm}$\text{$0.77$}$\vspace{0.1mm}} &  \thead{(0.2465, 0.34$\pi$)} & \thead{(0.0115, 1.19$\pi$)} &  \thead{(0.2137, 0.27$\pi$)} & \thead{(0.0175, 1.07$\pi$)} 
          \\  \hline

          \thead{\vspace{0.1mm}$\text{$0.87$}$\vspace{0.1mm}}  &  \thead{(0.2776, 0.34$\pi$)} & \thead{(0.0128, 1.19$\pi$)} &  \thead{(0.2403, 0.27$\pi$)} & \thead{(0.0192, 1.07$\pi$)} 
          \\  \hline
               \multicolumn{5}{|c|} {Bardeen}
          \\  \hline
          
           \thead{$\vspace{0.1mm}\text{$0.45$}$\vspace{0.1mm}}  &  \thead{(0.0456, 0.35$\pi$)} & \thead{(0.0025, 1.16$\pi$)} &  \thead{(0.0395, 0.28$\pi$)} & \thead{(0.0038, 1.05$\pi$)} 
          \\  \hline

          \thead{\vspace{0.1mm}$\text{$0.55$}$\vspace{0.1mm}} &  \thead{(0.0593, 0.35$\pi$)} & \thead{(0.0033, 1.16$\pi$)} &  \thead{(0.0514, 0.28$\pi$)} & \thead{(0.0050, 1.05$\pi$)} 
          \\  \hline

          \thead{\vspace{0.1mm}$\text{$0.7$}$\vspace{0.1mm}}  &  \thead{(0.0813, 0.35$\pi$)} & \thead{(0.0045, 1.16$\pi$)} &  \thead{(0.0703, 0.28$\pi$)} & \thead{(0.0067, 1.05$\pi$)} 
          \\  \hline
                            \multicolumn{5}{|c|} {Frolov}
          \\  \hline
          
           \thead{\vspace{0.1mm}$\text{$0.63$}$\vspace{0.1mm}} &  \thead{(0.0100, 0.35$\pi$)} & \thead{(0.0006, 1.15$\pi$)} &  \thead{(0.0086, 0.29$\pi$)} & \thead{(0.0009, 1.03$\pi$)} 
          \\  \hline

          \thead{\vspace{0.1mm}$\text{$0.73$}$\vspace{0.1mm}} &  \thead{(0.0139, 0.35$\pi$)} & \thead{(0.0008, 1.15$\pi$)} &  \thead{(0.0120, 0.29$\pi$)} & \thead{(0.0013, 1.03$\pi$)} 
          \\  \hline

          \thead{\vspace{0.1mm}$\text{$1.05$}$\vspace{0.1mm}} &  \thead{(0.0319, 0.35$\pi$)} & \thead{(0.0019, 1.15$\pi$)} &  \thead{(0.0275, 0.29$\pi$)} & \thead{(0.0029, 1.04$\pi$)} 
          \\  \hline

       \end{tabular}
      \caption{Maximum deviations of the polarized intensity and EVPA from the Schwarzschild reference for an observer inclination $i=20^\circ$ and the magnetic-field configurations specified in the table. For each regular geometry and value of $\ell$, we report $\max(\Delta I)/I_{\rm Sch}$ and $\max(\Delta{\rm EVPA})/{\rm EVPA}_{\rm Sch}$, together with the azimuthal angle $\phi$ at which each maximum occurs. The Schwarzschild reference corresponds to an emitting orbit at $r_{\rm source}=6m$, with the observer located at $r_{\rm obs}=10^4m$. The three values of $\ell$ for each solution represent from top to bottom a black-hole configuration, an ultracompact configuration, and a horizonless configuration without photon spheres.}
    \label{table:theta_20_1}
\end{table}

The four regular spacetimes considered in this work are controlled by a single regularization parameter $\ell$. As discussed in Section 2, the characteristic radii of the spacetimes exhibit a qualitatively similar evolution as $\ell$ varies describing a black hole for small values of $\ell$ and going through an ultracompact particle-like geometry and a particle-like geometry without light rings when $\ell$ grows. This allows us to compare configurations belonging to analogous geometric regimes across the four metrics. To make this comparison physically meaningful, we choose values of $\ell$
corresponding to the same qualitative spacetime character.

\begin{table}[t]
    \centering
      \begin{tabular}{|c|c|c|c|c|}
       \hline
       \thead{ $i = 20^{\circ}$ }& \multicolumn{2}{c|}{{$B = [0.5, 0.87, 0]$, $\chi=-120\degree$}} & \multicolumn{2}{c|}{{$B = [0, 0, 1]$, $\chi=-150\degree$}} 
          \\  \hline
          
         \thead{ $\ell$}   &  \thead{$\left(\frac{\text{max}\,\Delta \text{I}}{\text{I}_\text{Sch}} \,, \, \phi \, \right)$} & \thead{$\left(\frac{\text{max}\,\Delta \text{EVPA}}{\text{EVPA}_\text{Sch}} \,, \, \phi \, \right)$} &  \thead{$\left(\frac{\text{max}\,\Delta \text{I}}{\text{I}_\text{Sch}} \,, \, \phi \, \right)$} & \thead{$\left(\frac{\text{max}\,\Delta \text{EVPA}}{\text{EVPA}_\text{Sch}} \, , \, \phi \, \right)$} 
          \\  \hline
          \multicolumn{5}{|c|} {Hayward}
          \\  \hline
           \thead{$\vspace{0.1mm}\text{$0.75$}$\vspace{0.1mm}}  &  \thead{(0.0078, 0.23$\pi$)} & \thead{(0.0011, 0.96$\pi$)} &  \thead{(0.0089, 0.38$\pi$)} & \thead{(0.0073, 0.60$\pi$)} 
          \\  \hline

          \thead{\vspace{0.1mm}$\text{$0.85$}$\vspace{0.1mm}} &  \thead{(0.0100, 0.23$\pi$)} & \thead{(0.0014, 0.96$\pi$)} &  \thead{(0.0114, 0.38$\pi$)} & \thead{(0.0093, 0.60$\pi$)} 
          \\  \hline

          \thead{\vspace{0.1mm}$\text{$1.3$}$\vspace{0.1mm}}  &  \thead{(0.0232,  0.23$\pi$)} & \thead{(0.0033, 0.96$\pi$)} &  \thead{(0.0265, 0.38$\pi$)} & \thead{(0.0218, 0.60$\pi$)} 
          \\  \hline
                  \multicolumn{5}{|c|} {Simpson}
          \\  \hline
          
           \thead{$\vspace{0.1mm}\text{$0.70$}$\vspace{0.1mm}}  &  \thead{(0.1642, 0.2$\pi$)} & \thead{(0.0183, 0.99$\pi$)} &  \thead{(0.0652, 1.26$\pi$)} & \thead{(0.1308, 0.59$\pi$)} 
          \\  \hline

          \thead{\vspace{0.1mm}$\text{$0.77$}$\vspace{0.1mm}} &  \thead{(0.1799, 0.2$\pi$)} & \thead{(0.0199, 0.99$\pi$)} &  \thead{(0.0711, 1.26$\pi$)} & \thead{(0.1403, 0.59$\pi$)} 
          \\  \hline

          \thead{\vspace{0.1mm}$\text{$0.87$}$\vspace{0.1mm}}  &  \thead{(0.2021, 0.2$\pi$)} & \thead{(0.0219, 0.99$\pi$)} &  \thead{(0.0795, 1.26$\pi$)} & \thead{(0.1590, 0.59$\pi$)} 
          \\  \hline
              \multicolumn{5}{|c|} {Bardeen}
          \\  \hline
          
           \thead{$\vspace{0.1mm}\text{$0.45$}$\vspace{0.1mm}}  &  \thead{(0.0334, 0.22$\pi$)} & \thead{(0.0044, 0.97$\pi$)} &  \thead{(0.0128, 1.30$\pi$)} & \thead{(0.0297, 0.60$\pi$)} 
          \\  \hline

          \thead{\vspace{0.1mm}$\text{$0.55$}$\vspace{0.1mm}} &  \thead{(0.0431, 0.21$\pi$)} & \thead{(0.0056, 0.97$\pi$)} &  \thead{(0.0166, 1.30$\pi$)} & \thead{(0.0387,0.60$\pi$)} 
          \\  \hline

          \thead{\vspace{0.1mm}$\text{$0.7$}$\vspace{0.1mm}}  &  \thead{(0.0589, 0.21$\pi$)} & \thead{(0.0077, 0.97$\pi$)} &  \thead{(0.0226, 1.30$\pi$)} & \thead{(0.0495, 0.59$\pi$)} 
          \\  \hline
                            \multicolumn{5}{|c|} {Frolov}
          \\  \hline
          
           \thead{\vspace{0.1mm}$\text{$0.63$}$\vspace{0.1mm}} &  \thead{(0.0072, 0.22$\pi$)} & \thead{(0.0010, 0.96$\pi$)} &  \thead{(0.0082, 0.38$\pi$)} & \thead{(0.0065, 0.60$\pi$)} 
          \\  \hline

          \thead{\vspace{0.1mm}$\text{$0.73$}$\vspace{0.1mm}} &  \thead{(0.0100, 0.22$\pi$)} & \thead{(0.0014, 0.96$\pi$)} &  \thead{(0.0114, 0.38$\pi$)} & \thead{(0.0091, 0.60$\pi$)} 
          \\  \hline

          \thead{\vspace{0.1mm}$\text{$1.05$}$\vspace{0.1mm}} &  \thead{(0.0230, 0.22$\pi$)} & \thead{(0.0032, 0.96$\pi$)} &  \thead{(0.0262, 0.38$\pi$)} & \thead{(0.0210, 0.60$\pi$)} 
          \\  \hline

       \end{tabular}
      \caption{Maximum deviations of the polarized intensity and EVPA from the Schwarzschild reference for an observer inclination $i=20^\circ$ and the magnetic-field configurations specified in the table. For each regular geometry and value of $\ell$, we report $\max(\Delta I)/I_{\rm Sch}$ and $\max(\Delta{\rm EVPA})/{\rm EVPA}_{\rm Sch}$, together with the azimuthal angle $\phi$ at which each maximum occurs. The Schwarzschild reference corresponds to an emitting orbit at $r_{\rm source}=6m$, with the observer located at $r_{\rm obs}=10^4m$. The three values of $\ell$ for each solution represent from top to bottom a black-hole configuration, an ultracompact configuration, and a horizonless configuration without photon spheres.}
    \label{table:theta_20_2}
\end{table}

For each regular metric, we select three representative values of the parameter $\ell$, chosen to showcase the three qualitatively distinct regimes. The first lies in the black-hole regime, where event horizons are present. The second describes a horizonless configuration that still possesses photon spheres. The third corresponds to a spacetime in which neither horizons nor photon spheres are present, with
$\ell$ sufficiently large that the marginally bound orbit (MBO) and the marginally stable circular orbit (MSCO) no longer exist (see Table \ref{crit_str}).

\begin{figure}[t!]
    \centering
    \includegraphics[width=14.2cm]{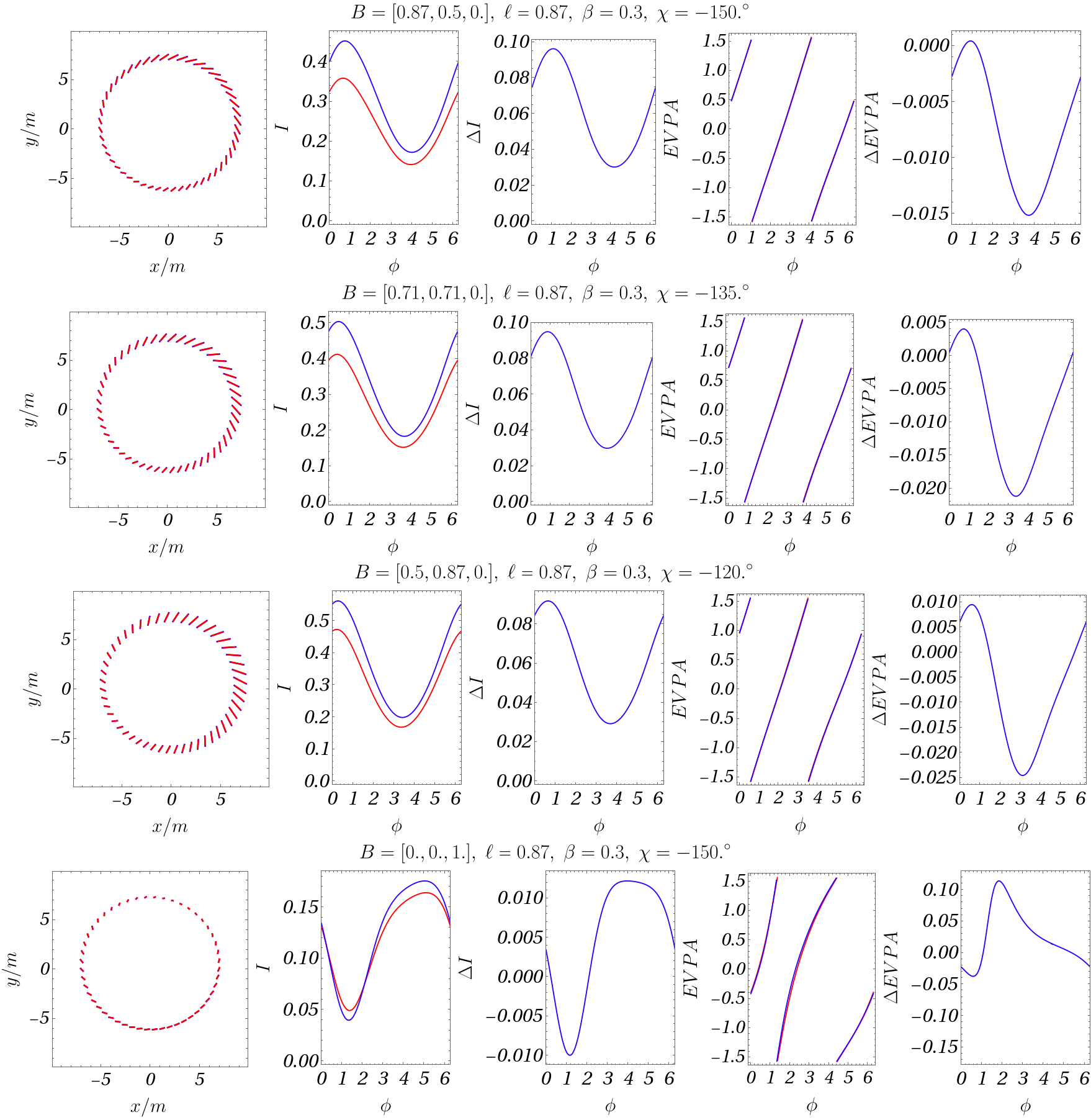}
    \caption{Linear polarization for the Simpson geometry with $\ell=0.87$ (blue), compared with the Schwarzschild reference (red), for an observer inclination $i=20^\circ$. The Schwarzschild emitting orbit is located at
$r_{\rm source}=6m$, and the observer at $r_{\rm obs}=10^4m$. The panels show the polarized intensity $I$, the EVPA, and the corresponding differences $\Delta I$ and $\Delta{\rm EVPA}$ relative to the Schwarzschild case.}
    \label{fig:Simpson_vs_Shwarzschild_i=20_l=0.87}
\end{figure}

We compare the polarized intensity and EVPA  for four representative magnetic field configurations and  observer inclination $i=20^{\circ}$. We present our results for the magnetic fields  $B=\left[\hat{B}^r, \hat{B}^{\phi}, \hat{B}^{\theta}\right] = \left[0.87,0.5,0\right]$ and $B=\left[0.71,0.71,0\right]$ in Table~\ref{table:theta_20_1}, whereas  $B=\left[0.5,0.87,0\right]$ and $B=\left[0,0,1\right]$ are shown in Table~\ref{table:theta_20_2}. For each configuration we evaluate the deviation of the regular geometry (RG) from the Schwarzschild black hole by computing the relative intensity $\Delta I = I_{\rm RG}-I_{\rm Sch}$ and the relative direction angle $\Delta_{\rm EVPA} = {\rm EVPA}_{\rm RG} - {\rm EVPA}_{\rm Sch}$ at each point of the circular orbit image. Then, we present in the tables the maximum deviation of the polarization intensity from the Schwarzschild black hole which is reached along the image  normalized by the polarization intensity for the Schwarzschild black hole at the corresponding point $\max\,\Delta {\rm I}/{\rm I}_{\rm Sch}$. We adopt a similar measure for the deviation of the EVPA along the image calculating the maximum deviation from the Schwarzschild black hole reached in the image normalized by the EVPA for the Schwarzschild black hole at this point. $\max\,\Delta{\rm EVPA}/{\rm EVPA}_{\rm Sch}$.

We observe that for all the geometries the deviation from the Schwarzschild black hole increases with growing regularization parameter $\ell$. Thus, particle-like geometries without light rings require the lowest resolution to be distinguished from the Schwarzschild black hole. In terms of magnetic field configurations these particle-like geometries reach maximum deviation in the polarization intensity for the equatorial magnetic field with a dominating radial component $B= \left[0.87,0.5,0\right]$ and maximum deviation in the EVPA for the vertical magnetic field $B= \left[0,0,1\right]$ . Regular geometries with Minkowski core represented by the Simpson spacetime show larger deviations with $\max\,\Delta {\rm I}/{\rm I}_{\rm Sch}$  reaching $\approx 21\%$ and $\max\,\Delta{\rm EVPA}/{\rm EVPA}_{\rm Sch}$ reaching $\approx 16\%$ for the particle-like solutions. The geometries with de Sitter core demonstrate maximum deviations from the Schwarzschild black hole in the ranges $\max\,\Delta {\rm I}/{\rm I}_{\rm Sch} \approx 3\% - 8\%$ and  $\max\,\Delta{\rm EVPA}/{\rm EVPA}_{\rm Sch} \approx 2\% - 5\%$  for the same class of solutions.

Representative examples of these comparisons on the observer's image plane are shown  in Fig.~\ref{fig:Simpson_vs_Shwarzschild_i=20_l=0.87} for the particle-like solutions without light rings in the Simpson geometry where the deviations are most pronounced. Further polarized images for regular spacetimes with de Sitter core are provided in Appendix A (see Fig. \ref{fig:Hayward_vs_Shwarzschild_i=20_l=1.3}). For completeness  we also examine in the Appendix the behavior of the polarization properties for higher inclination angles. We choose suitable inclinations for which the circular orbits still possess a single direct image and perform a quantitative comparison of the polarization intensity and EVPA with the Schwarzschild black hole similar to the analysis performed in this section (see Tables $\ref{table:theta_70_1}-\ref{table:theta_70_2}$ and Figs. $\ref{fig:Hayward_vs_Shwarzschild_i=70_l=1.3}-\ref{fig:Simpson_vs_Shwarzschild_i=70_l=0.87}$).

\section{Linear polarization for particle-like solutions at high inclinations}

In this section we examine the polarization from the inner direct images of equatorial circular sources which are characteristic for high inclination of the observer. As we discussed in Section 4 the formation of these images depends on a set of parameters. If we consider an asymptotic observer they include the spacetime regularization parameter, the radial location of the emitting source, and the inclination angle.

\begin{figure}[t!]
    \centering    
    \includegraphics[width=14cm]{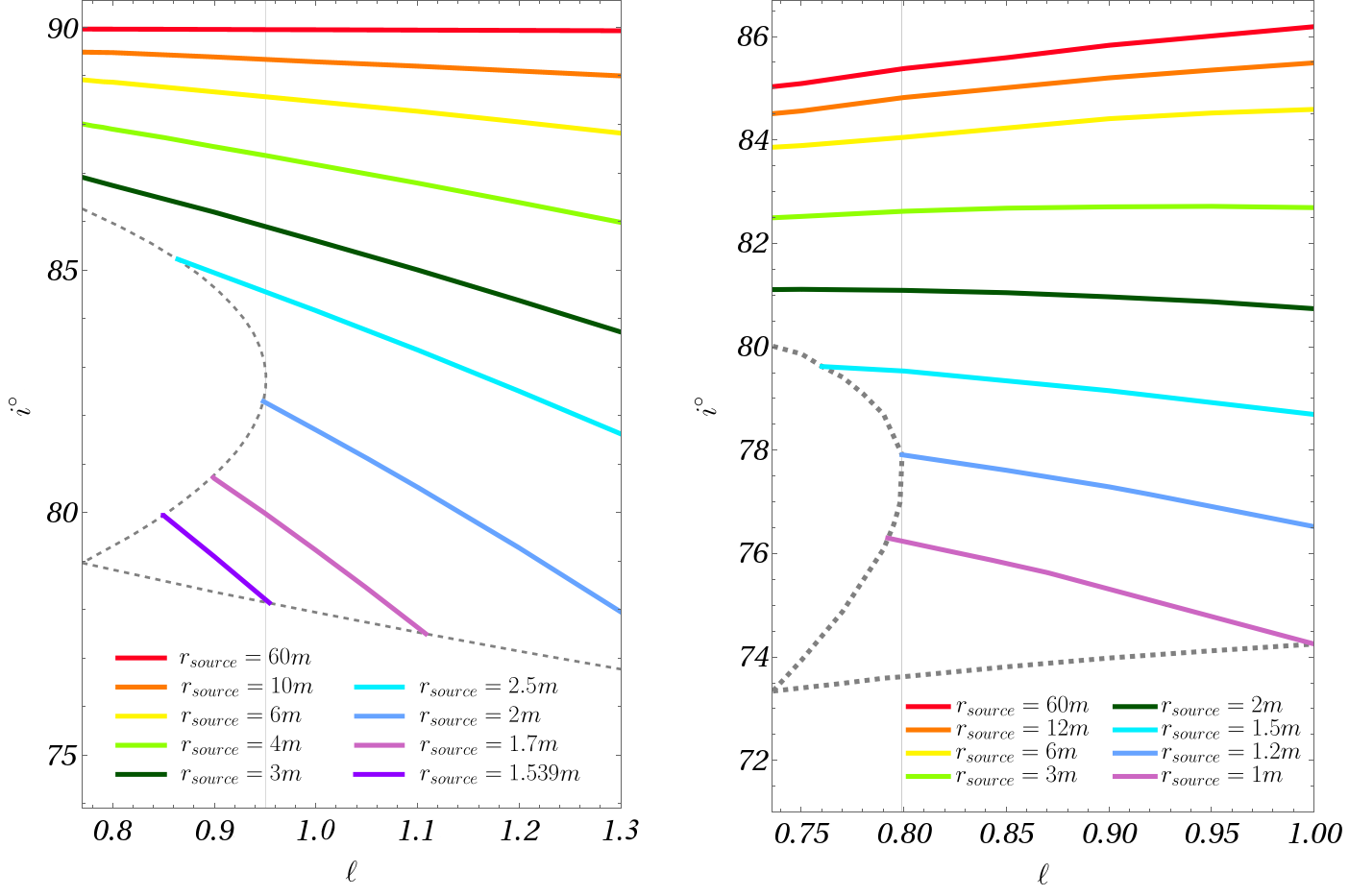}
   \caption{Parameter space for the existence of the inner direct images in the Hayward spacetime (left) and the Simpson spacetime (right). The vertical axis gives the critical inclination for the formation of inner images, while the horizontal axis represents the regularization parameter $\ell$. The different curves correspond to different radial positions of the emitting orbit $r_{source}$ with color coding given by the legend.  As $r_{source}$ increases, the curves approach a horizontal asymptote which can be approximated with the red curve. We denote by dashed lines the location of the stable and unstable photon spheres and the location of the ICOs. The vertical grey line corresponds to the regularization parameter $\ell_{ph}$ where the two photon spheres merge and provides the limit between ultracompact geometries and geometries without light rings. The observer is located at $r_{obs}=10^{4}\,m$.}
    \label{fig:Hayward_Simpson_inner_image}
\end{figure}

Two effects take place which determine the observability of the inner images. On the one hand, the regularization parameter and the source location determine the shape of the deflection curve (see Figs. \ref{fig:Image_part}-\ref{fig:Image_Simpson}) and in particular the location of the local minimum at small impact parameters which is responsible for the formation of the inner direct images. On the other hand, the inclination angle controls the width of the observational interval. Observable images correspond to intersections of the deflection curve with the observational interval. Too low inclinations lead to narrower observational intervals for the direct images which fail to encompass the local minimum. Consequently,  an inner direct images is not observed. The two effects interplay in a complicated manner. Therefore, we examined numerically the parameter space where the inner direct images exist presenting our analysis in Fig. $\ref{fig:Hayward_Simpson_inner_image}$. We show two representative spacetimes with de Sitter and Minkowski cores, while further examples are presented in Appendix B in Fig. $\ref{fig:Bardeen_Frolov_inner_image}$.

Figs. $\ref{fig:Hayward_Simpson_inner_image}$ and $\ref{fig:Bardeen_Frolov_inner_image}$  represent the critical inclination necessary for the inner direct image to become observable as a function of the regularization parameter $\ell$ and the location of the emitting source $r_{source}$. For each set of values ($\ell$, $r_{source}$) the inner direct image is observable for inclinations higher than the critical inclination corresponding to ($\ell$, $r_{source}$) according to the parameter space diagram. We observe the following correlations. The critical inclination increases with increasing radial distance of the emitting source. For larger distances of the emitting source, e.g $r_{source}\gtrsim 6m$, the critical inclination depends weakly on the regularization parameter. It retains very high values as $i\gtrsim 88^{\circ}$ for spacetimes with de Sitter cores and $i\gtrsim 83^{\circ}$ for the Simpson spacetime with Minkowski core. Thus, the inner direct images are observable only for nearly edge-on accretion disks.

\begin{figure}[t!]
    \centering
    \includegraphics[width=14.2cm]{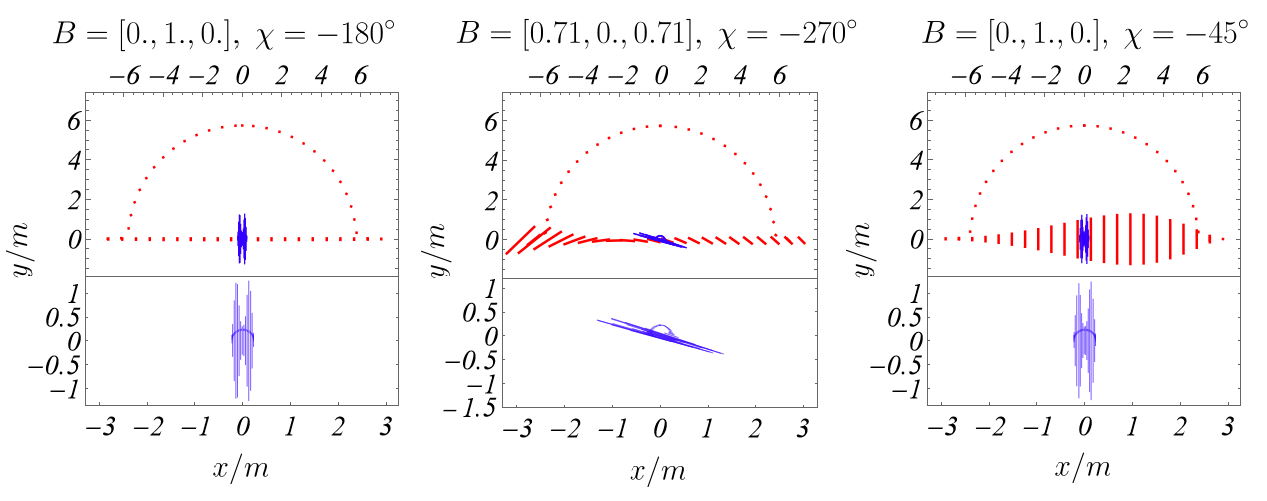}
    \caption{{\small Examples of polarized images for the ultracompact Simpson geometry with regularization parameter  $\ell=0.79$ Varying  the magnetic field and the fluid velocity direction we observe different intensity and polarization pattern for the inner (blue) and outer (red) images of the circular orbits. The emitting ring is located at $r_{source}=6m$ and the intensity profiles of the two images are normalized to the maximum intensity of the inner image. The observer is located at $r_{\rm obs}=10^4m$ and $i=89.9^\circ$.} }
    \label{fig:3examples}
\end{figure}

For smaller distances of the emitting source, e.g. $r_{source}\lesssim 3m$, lower critical inclination become possible, as the inclination depends more significantly on the regularization parameter and further decreases with growing $\ell$. However, there exists a lower limit for the possible critical inclination in a particular spacetime. On the one hand, it is determined by the lower limit of existence of the timelike circular orbits which is set by the ICOs (see Fig. \ref{Hayward_str}). On the other hand, we need to impose some restriction on the natural values of the regularization parameter $\ell$. In our analysis we consider an upper limit $l\sim 1$ which provides another constraint on the lower limit of critical inclination. Thus, we obtain that emitting sources with  $r_{source}\lesssim 3m$ produce inner direct images in the Hayward spacetime for regularization parameters $77^\circ\lesssim i \lesssim 87^{\circ}$ with similar values for the Bardeen and Frolov spacetimes (see Appendix B). In the Simpson spacetime the critical inclination becomes lower for the same range of emitting radii belonging to the approximate interval $73^\circ\lesssim i\lesssim 83^{\circ}$.

In general we observe that the Simpson spacetime leads to smaller critical inclinations that the spacetimes with de Sitter core for any regularization parameter and emitting radius. In addition, particle-like solutions with no light rings possess smaller critical inclinations than ultracompact particle-like solutions for any of the geometries since they arise  at higher values of the regularization parameter $\ell$. Thus, they favor the formation of an inner direct image of the circular orbits.

\subsection{Intensity of the polarized radiation from the inner images}

\begin{figure}[t!]
    \centering
    \includegraphics[width=14.2cm]{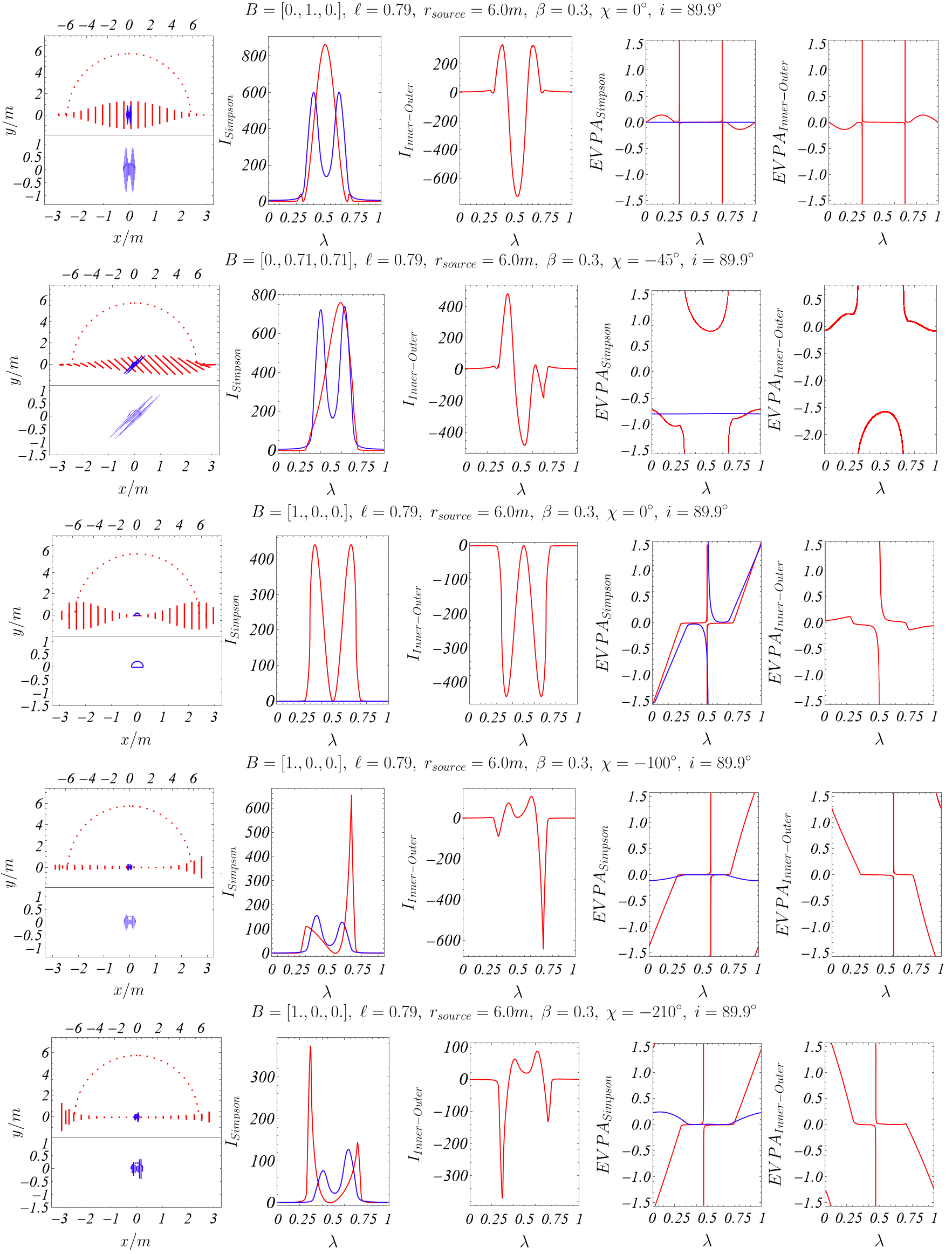}
    \caption{{\small Examples of different magnetic field and emitting medium configurations. The polarized intensity of the outer image is greater than or equal to that of the inner image. In each row, the intensity profiles of the two images are normalized to the maximum intensity of the outer image, given by the highest peak in the second panel. In the leftmost panels, the lengths of the polarization ticks are proportional to the normalized polarized intensity. The observer is located at $r_{\rm obs}=10^4m$ and $i=89.9^\circ$.}}
    \label{fig:EVPApanel1}
\end{figure}

\begin{figure}[t!]
    \centering
    \includegraphics[width=14.2cm]{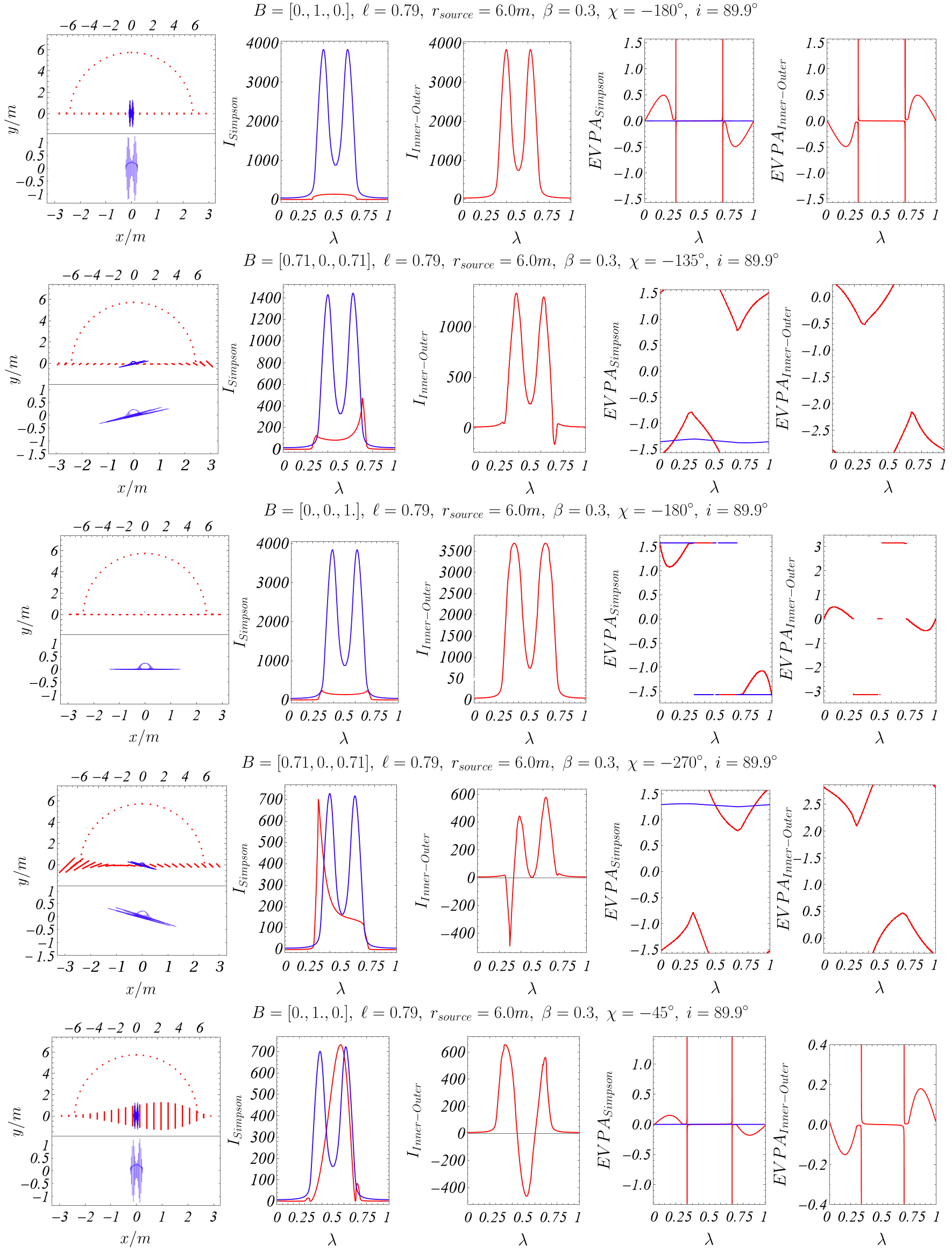}
    \caption{{\small Examples of different magnetic field and emitting medium configurations. The polarized intensity of the inner image is greater than or equal to that of the outer image. In each row, the intensity profiles of the two images are normalized to the maximum intensity of the inner image, given by the highest peak in the second panel. In the leftmost panels, the lengths of the polarization ticks are proportional to the normalized polarized intensity. The observer is located at $r_{\rm obs}=10^4m$ and $i=89.9^\circ$.}}
    \label{fig:EVPApanel2}
\end{figure}

The direct inner images radiate with very different intensity depending on the magnetic field configuration and the plasma velocity. As a measure for the relative intensity of the inner images we adopt their comparison with the outer images of the same circular orbits which are observed at  higher impact parameters. Our simulations show that for some plasma configurations the intensity of the inner images is below $1\%$ of the intensity of the outer images while for others it becomes with three orders of magnitude larger (see Fig. \ref{fig:3examples} and Figs. \ref{fig:EVPApanel1}-\ref{fig:EVPApanel2})

In order to get intuition which plasma configurations trigger such an amplification of the polarized emission we study numerically the parametric space which describes the plasma environment. For convenience we consider a magnetic field which is confined to a single plane. Then, we can parameterize the magnetic field direction by the angle $\eta$ between its two components.  On the other hand, the plasma velocity direction is parameterized by the angle $\chi$ (see Eq. $\ref{eq:betaboost}$ ) where we assume that the plasma motion is confined to the equatorial plane. Thus, we can explore the intensity of the polarized radiation from the inner images as a function of the two angles $(\eta, \chi)$. We repeat this analysis for all the three coordinate planes in which the magnetic field may lie. In this way, we can obtain intuition for the regions in the parametric space which induce amplification of the inner image intensity for a general three-dimensional magnetic field configuration.

\begin{figure}[t!]
    \centering
    \includegraphics[width=15.2cm]{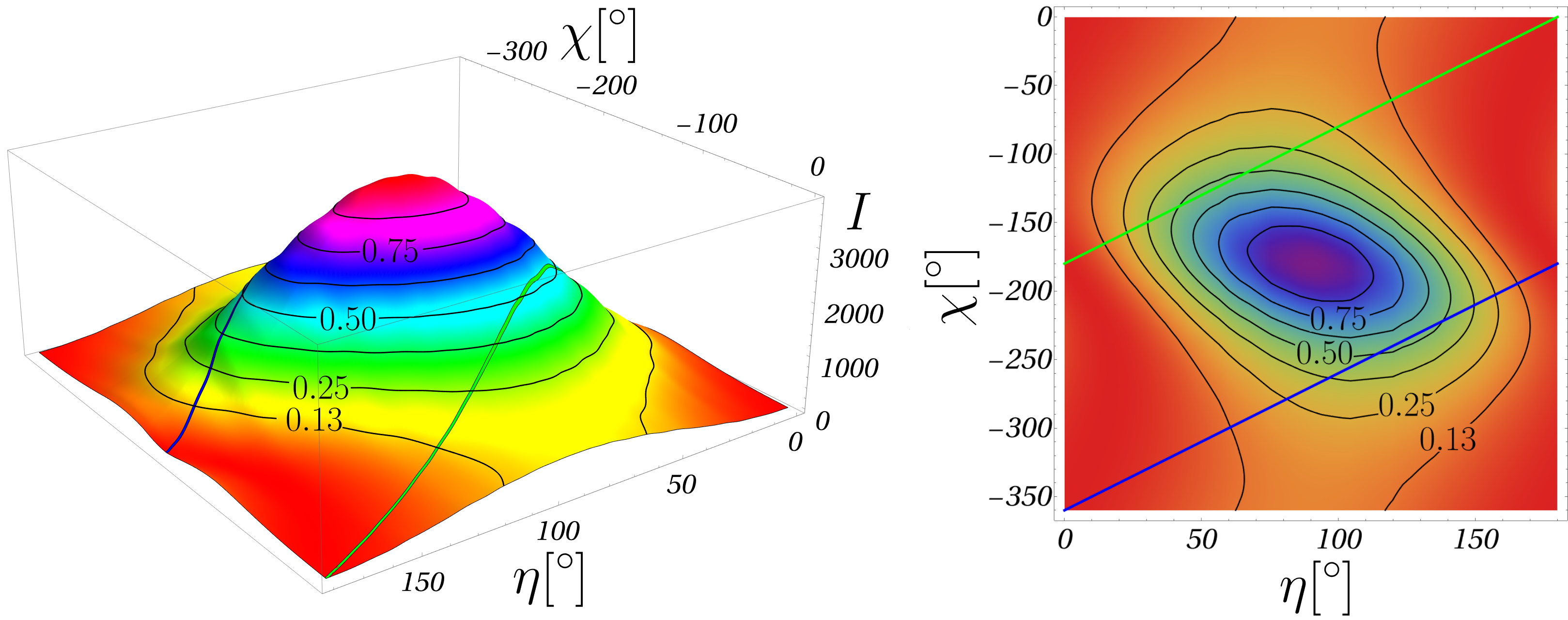}
    \caption{Polarization intensity from the equatorial
    emitting ring at $r_{source}=6m$ in the Simpson spacetime as a function of the
    plasma velocity direction $\chi$ and the magnetic field orientation
    $\eta$. The magnetic field is confined to the equatorial
    $(\hat r,\hat\phi)$ plane. The blue and green curves correspond to
    $\eta=\chi$ and $\eta=\chi+\pi$, respectively. The maximum intensity
    displayed in the full parameter space is
    $I_{\rm max}^{\rm global}=3839.55$. The maxima attained along the blue
    and green curves are discussed in the main text. The observer is located
    at $r_{\rm obs}=10^4m$ with inclination $i=89.9^\circ$, and the metric
    parameter is $\ell=0.79$.}
    \label{fig:r_to_phi_Simpson}
\end{figure}

\begin{figure}[t!]
    \centering
    \includegraphics[width=15.2cm]{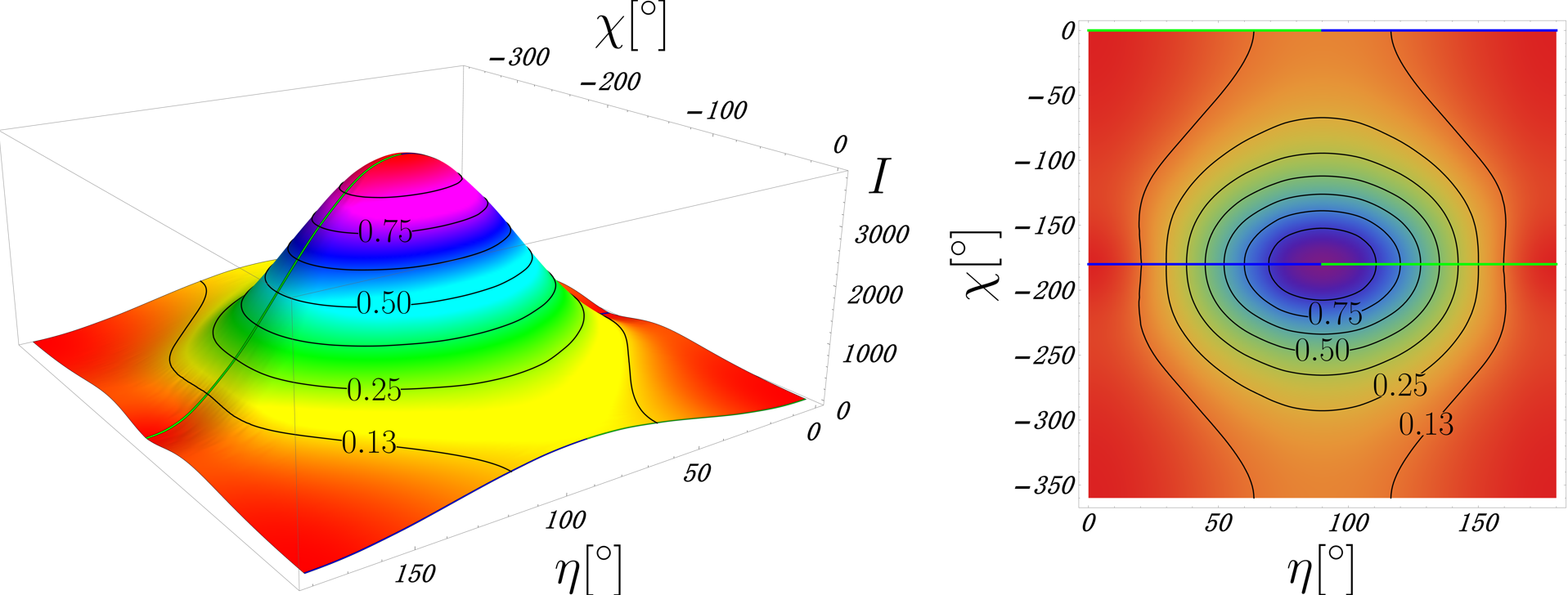}
    \caption{Polarization intensity from the equatorial
    emitting ring at $r_{source}=6m$ in the Simpson spacetime as a function of the
    plasma velocity direction $\chi$ and the magnetic field orientation
    $\eta$. The magnetic field is confined to the
    $(\hat r,\hat\theta)$ plane. The blue and green curves correspond to
    configurations with fluid velocity parallel or anti-parallel to the equatorial component of the magnetic field, respectively. The maximum intensity displayed in the full parameter space is
    $I_{\rm max}^{\rm global}= 3841.75$. The maxima attained
    along these two configurations are discussed in the main text.
    The observer is located at $r_{\rm obs}=10^4m$ with inclination
    $i=89.9^\circ$, and the metric parameter is $\ell=0.79$.}
    \label{fig:r_to_theta_Simpson}
\end{figure}

\begin{figure}[t!]
    \centering
    \includegraphics[width=15.2cm]{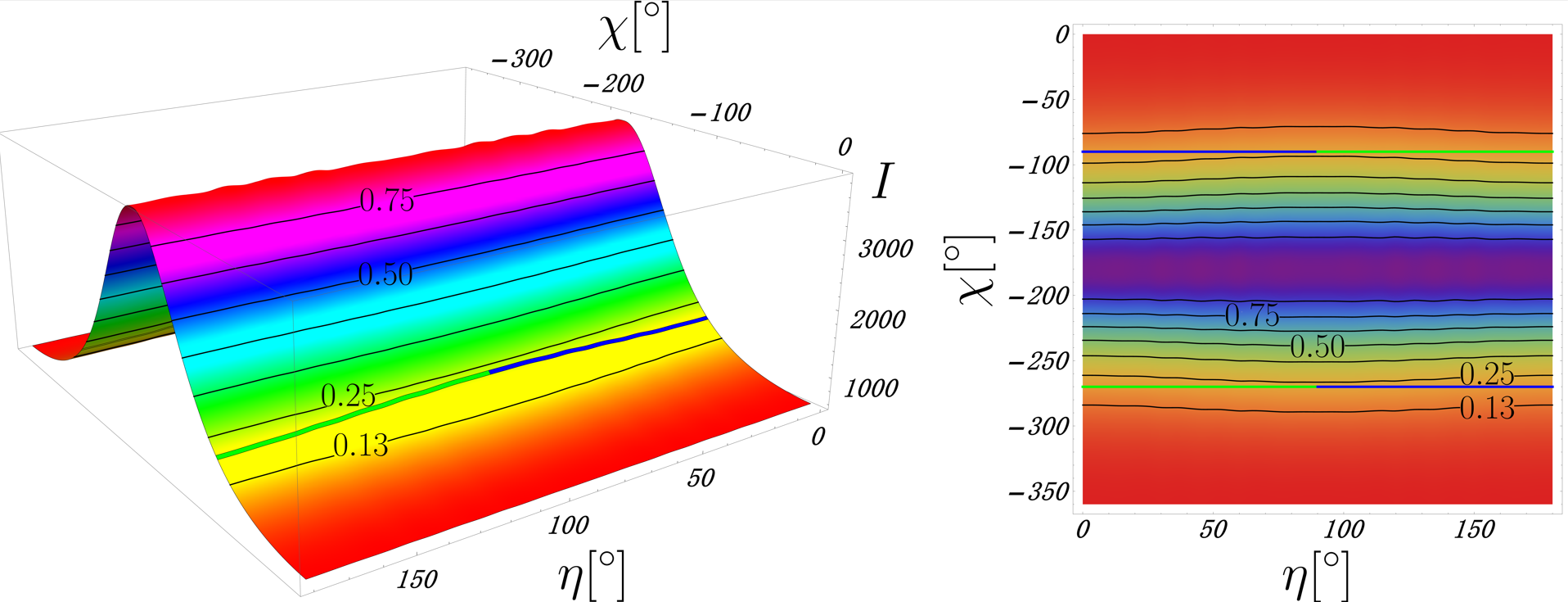}
    \caption{Polarization intensity from the equatorial
    emitting ring at $r_{source}=6m$ in the Simpson spacetime as a function of the plasma velocity direction $\chi$ and the magnetic field orientation $\eta$. The magnetic field is confined to the
    $(\hat\phi,\hat\theta)$ plane. The blue and green curves correspond to
    configurations with fluid velocity parallel or anti-parallel to the equatorial component of the magnetic field, respectively. The maximum intensity displayed in the full parameter space is
    $I_{\rm max}^{\rm global}= 3871.67$. The maxima attained
    along these two configurations are discussed in the main text.
    The observer is located at $r_{\rm obs}=10^4m$ with inclination
    $i=89.9^\circ$, and the metric parameter is $\ell=0.79$.}
    \label{fig:phi_to_theta_Simpson}
\end{figure}

Since we aim at exploring the influence of the plasma environment on the polarization intensity we perform our analysis for a  fixed spacetime geometry with fixed boundary conditions for the light propagation. As a representative example we select the ultracompact particle-like Simpson solution with regularization parameter $\ell =0.79$. The emitting source is located at radial distance $r_{source} = 6m$ while the observer is at the asymptotic infinity with an inclination angle $i=89.9^{\circ}$.

In Fig.~\ref{fig:r_to_phi_Simpson} we explore the polarization intensity for a magnetic field confined to the equatorial plane, i.e. $\vec{B}=\left[\hat{B}^{r},\hat{B}^{\phi},0\right]$. The magnetic field $\vec{B}$ is parameterized by the angles $\eta$ and $\chi$ as
\begin{eqnarray}\label{B_r_phi}
\vec{B}=B\left(
\cos\eta\,e_{(r)}
+\sin\eta\,e_{(\phi)}
\right).
\end{eqnarray}

We observe that radial magnetic fields inhibit the polarization intensity while azimuthal ones lead to amplification. On the other hand, radial outward fluid velocity ($\chi=0$) leads to a low intensity and the intensity increases continuously as the velocity vector rotates toward radial inward direction ($\chi=-\pi$). The two effect interplay in a complicated way and in general for a fixed magnetic field the intensity maximum does not correspond to the purely inward radial direction of the fluid velocity but is shifted from the line $\chi=-\pi$. The global maximum of the inner image intensity occurs at $\eta=\pi/2$, corresponding to a purely azimuthal magnetic field, and $\chi=-\pi$, corresponding to a radial inward plasma motion.

In our analysis we regard the angles $\eta$ and $\chi$ as independent parameters. However, in many models of the accretion flow the magnetic field determines the direction of the fluid velocity. For equatorial magnetic fields they are most naturally related as $\eta=\chi$ or $\eta=\chi+\pi$ while for a general configuration the fluid velocity tends to be parallel or anti-parallel to the equatorial component of the magnetic field \cite{Narayan:2021}. We highlight these configurations denoting them with blue and green curves, respectively. Along these curves, the largest intensity is again obtained near $\eta=\pi/2$. The corresponding maximum polarized intensities are $I_{\rm max}^{(\eta=\chi)}=2055.91$ and $I_{\rm max}^{(\eta=\chi+\pi)}=2056.27$, corresponding to approximately
$53.55\%$ and $53.56\%$ of the global maximum $I_{\rm max}^{\rm global}=3839.55$.

Next we consider a magnetic field in the $(\hat r,\hat\theta)$ plane $\vec{B}=\left[\hat{B}^{r},0,\hat{B}^{\theta}\right]$ with the corresponding results shown in Fig.~\ref{fig:r_to_theta_Simpson}. 
The magnetic field $\vec{B}$ is parameterized by the angle $\eta$ and $\chi$ as
\begin{eqnarray}\label{B_r_theta}
\vec{B}=B\left(
\cos\eta\,e_{(r)}
+\sin\eta\,e_{(\theta)}
\right).
\end{eqnarray}
We observe again that radial magnetic fields inhibit the polarization intensity while polar ones lead to amplification. The intensity is lowest for fluid velocity in radial outward direction and highest for radial inward direction. In this case the two effects are detached and for a general fixed magnetic field the maximum intensity is reached at $\chi=-\pi$. The global intensity maximum is obtained for $\eta=\pi/2$, corresponding to a polar magnetic field. The physically motivated configurations with fluid velocity parallel or anti-parallel to the equatorial component of the magnetic field are indicated by the blue and green curves, respectively. They exhibit their largest intensity at the global maximum.

Finally, we consider a magnetic field in the $(\hat\phi,\hat\theta)$ plane $\vec{B}=\left[0,\hat{B}^{\phi},\hat{B}^{\theta}\right]$ illustrated in  in Fig.~\ref{fig:phi_to_theta_Simpson}.The magnetic field is parameterized by the angle $\eta$ as
\begin{eqnarray}\label{B_phi_theta}
\vec{B}=B\left(
\cos\eta\,e_{(\phi)}
+\sin\eta\,e_{(\theta)}
\right).
\end{eqnarray}
In contrast to the previous two cases, the intensity distribution develops a pronounced band-like structure in the $(\eta,\chi)$ parameter space and shows a very weak dependence on the magnetic field direction $\eta$. The polarized intensity is governed by the fluid velocity direction as we observe amplification for inward radial velocity and inhibition for outward radial plasma motion for any magnetic field configuration. For the configurations with fluid velocity parallel or anti-parallel to the equatorial component of the magnetic field, the peak polarized intensity is found to be $I_{\rm max}=1361.24$. Relative to the unconstrained maximum in the full $(\eta,\chi)$ parameter space, this peak corresponds to approximately $64.85\%$.

In summary we observe that the presence of a radial component in the magnetic field inhibits the polarization intensity of the inner direct images, while polar and azimuthal components amplify it at a comparable rate. The direction of the fluid velocity has an equally important impact and for any magnetic field the maximum intensity is reached for an approximately inward radial fluid velocity while the minimum intensity correspond to outward radial motion. Since the magnetic field and the plasma velocity directions are not expected to be constant in a realistic accretion disk, the strong dependence of the polarization intensity on these parameters may lead to interesting physical effects. Fluctuations in the plasma environment may lead to flare-ups in the inner image intensity when amplifying plasma configurations are reached. The presence of such flare-ups in the polarization intensity  may represent a specific observational signature for this class of regular particle-like geometries.

In our analysis in Figs. $\ref{fig:r_to_phi_Simpson}$-$\ref{fig:phi_to_theta_Simpson}$ we explored the magnitude of the maximum intensity reached in the polarized inner direct image. On the other hand, its location on the image of the circular orbit also depends on the magnetic field and plasma velocity directions. Depending on the particular plasma configuration the maximum intensity may be located on the left -or right-hand side of the image. There exist further configurations where we observe two local maxima with equal magnitude. In Figs. $\ref{fig:EVPApanel1}$-$\ref{fig:EVPApanel2}$ we present a gallery of images with different polarization patterns. We have organized the images so that Fig. $\ref{fig:EVPApanel1}$ shows configurations where the intensity of the inner direct image is low and the outer image dominates. On the other hand, Fig. \ref{fig:EVPApanel2}  highlights configurations with very high intensity of the inner direct image which exceeds the intensity of the outer image with several orders of magnitude. For better visualization purposes we represent the polarized images in terms of a parameter $\lambda$ which deviates from the polar angle $\phi$. It is defined as the length of the curve reached along the  image of the circular orbit on the observer's sky normalized by the full length of the orbit's image. The parameter $\lambda$ is measured with respect to the positive vertical axis in counter-clockwise direction, starting from the top of each image i.e. the axis  $(x=0, y>0)$ denotes the location of $\lambda=0$.

\subsection{EVPA of the polarized radiation from the inner images}

In this section we examine the behavior of the EVPA of the polarized radiation from the inner direct images. We perform a similar analysis as for the intensity in the previous subsection considering magnetic field  confined to a single plane and plasma velocity in the equatorial plane. This allows us to study the variation of the EVPA as a function of two angles $\eta$ and $\chi$ which determine the orientation of the magnetic field and the plasma velocity in the corresponding local basis. Performing this analysis for all the coordinate planes we can get intuition about the variation of the EVPA for a general three dimensional configuration of the plasma environment. As in our study of the polarization intensity we  select as an representative example the ultracompact particle-like Simpson solution with regularization parameter $\ell =0.79$. The emitting source is located at the radial distance $r_{source} = 6m$ while the observer is at the asymptotic infinity with an inclination angle $i=89.9^{\circ}$.

\begin{figure}[t!]
    		\setlength{\tabcolsep}{0pt}{\small\tt
		\begin{tabular}{ccc}
           \includegraphics[width=0.32\textwidth]{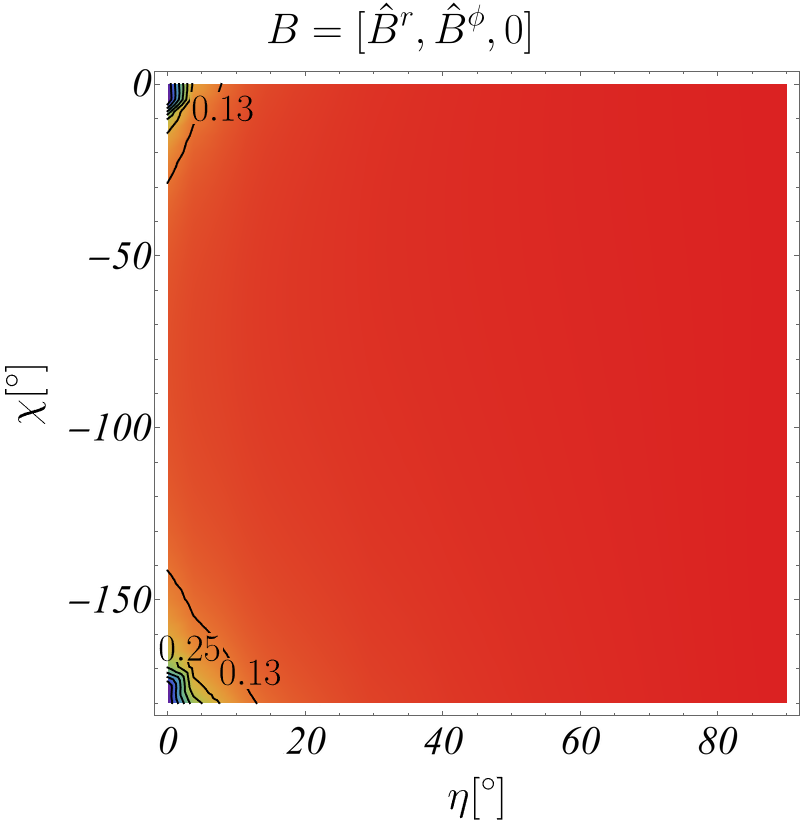}
		   \includegraphics[width=0.32\textwidth]{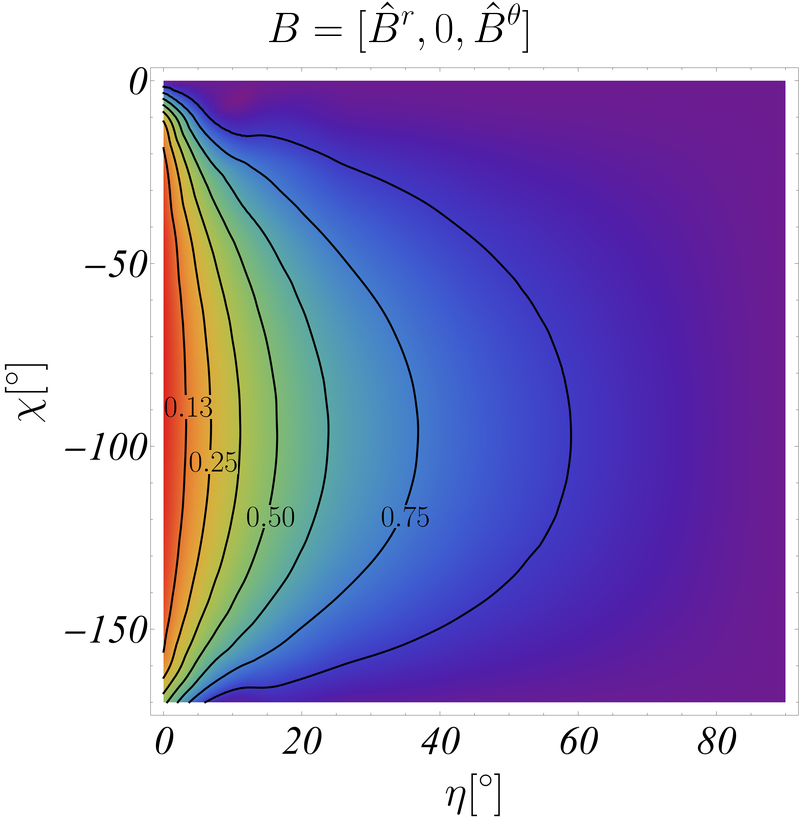} \includegraphics[width=0.32\textwidth]{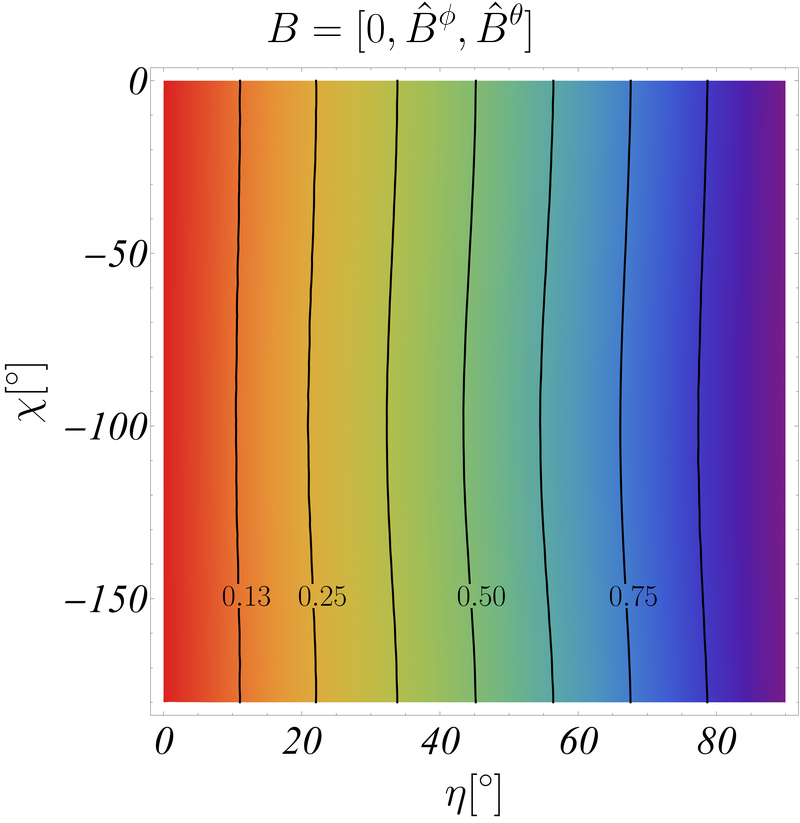}\\
          \hspace{0.8cm} a)  \hspace{4.3cm} b)   \hspace{4.3cm} c)
        \end{tabular}}
\caption{{Contour plots of the EVPA for the polarized emission from an emitting ring at $r_{source} = 6m$ in the Simpson spacetime as a function of the plasma velocity direction $\chi$ and the magnetic field orientation $\eta$. We analyze only the polarization from the inner direct image of the circular source. Three magnetic field configurations are presented where the magnetic field is confined to a single coordinate plane. The EVPA is normalized by its maximum absolute value $\pi/2$. The observer is located
at $r_{\rm obs}=10^4m$ with inclination $i=89.9^\circ$, and the regularization parameter is $\ell=0.79$.}
\label{fig:EVPA_contour}}
\end{figure}

\begin{figure}[t!]
    		\setlength{\tabcolsep}{0pt}{\small\tt
		\begin{tabular}{c}      
           \includegraphics[width=\textwidth]{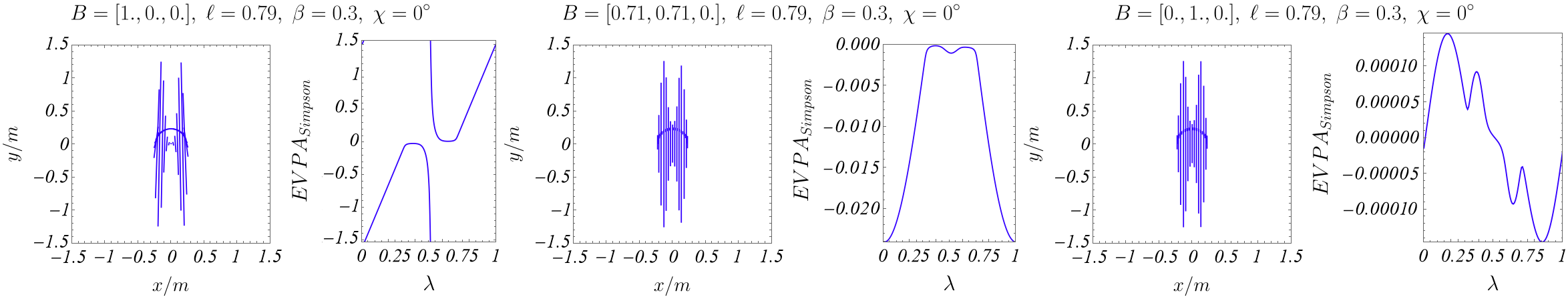}\\
		   \includegraphics[width=\textwidth]{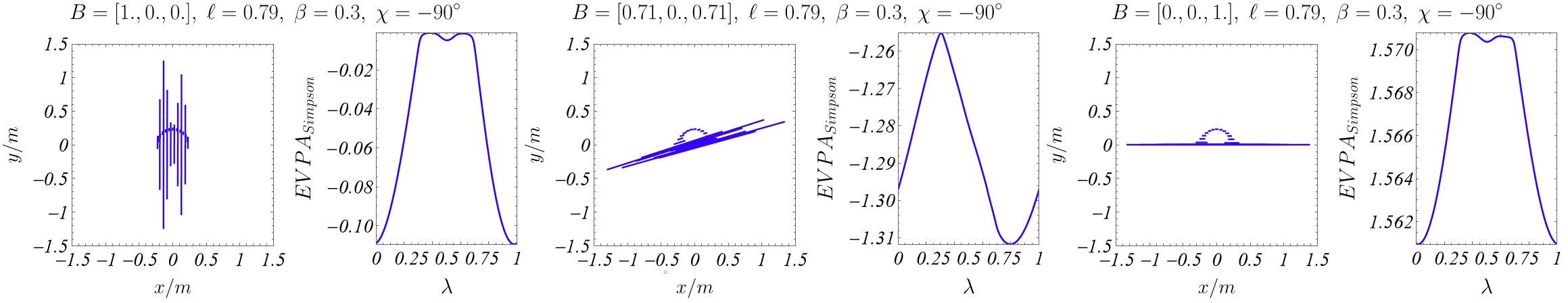} \\ \includegraphics[width=\textwidth]{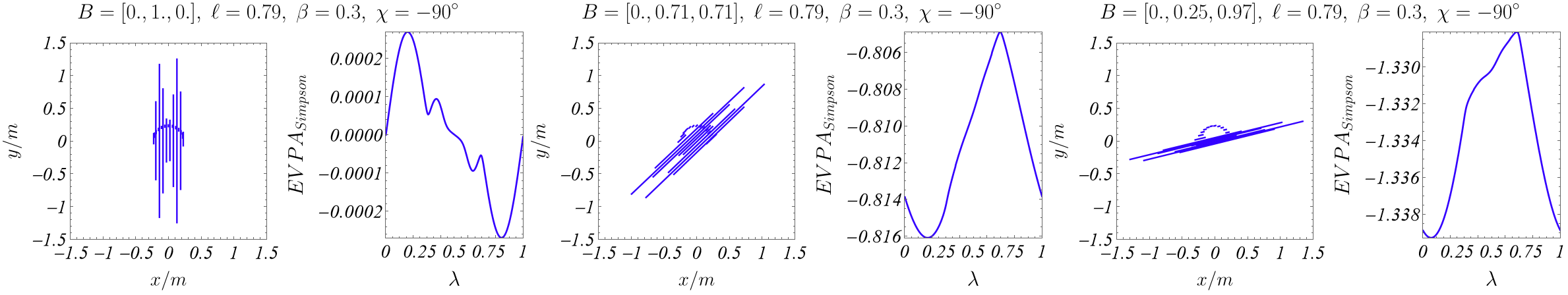}     
        \end{tabular}}
\caption{{Behavior of the EVPA for the polarized emission from an emitting ring at $r_{source} = 6m$ in the Simpson spacetime as the magnetic field direction varies. We show only the inner direct image of the circular source. In each row we present magnetic field configurations in which the magnetic is confined to a single coordinate plane. The observer is located at $r_{\rm obs}=10^4m$ with inclination $i=89.9^\circ$, and the regularization parameter is $\ell=0.79$.}
\label{fig:EVPA_image}}
\end{figure}

In Figs. \ref{fig:EVPA_contour}a -\ref{fig:EVPA_image} we examine the orientation of the EVPA for magnetic field and plasma velocity confined to the $(\hat r, \hat\phi)$-plane. The magnetic field has components $\vec{B}=\left[\hat{B}^{r},\hat{B}^{\phi}, 0\right]$ and it is represented in the local basis according to Eq. ($\ref{B_r_phi}$) while the plasma velocity is parametrized by the angle $\chi$ according to Eq. ($\ref{eq:betaboost}$). The variation of the EVPA in the $(\eta, \chi)$-parameter space is analyzed in the contour plot in Fig. \ref{fig:EVPA_contour}a  where we represent the absolute value of the maximum EVPA reached in the inner direct image for each plasma configuration normalized by the upper limit for the EVPA, i.e.  $|\frac{\text{max EVPA}}{\pi/2}|$. We observe that the maximum EVPA varies strongly only for configurations with an almost radial magnetic field and  plasma velocity direction close to $\chi=0$ and $\chi=-\pi$. For the predominant part of the parametric space the variation in the EVPA is very weak leading to an almost vertical direction of the polarization vector around the images. We illustrate this behavior in Fig. \ref{fig:EVPA_image} where we present the EVPA distribution for three equatorial magnetic fields and a constant plasma velocity direction $\chi=0$.

Next we consider magnetic field and plasma velocity lying in the $(\hat r, \hat\theta)$-plane with components of the magnetic field. $\vec{B}=\left[\hat{B}^{r},0,\hat{B}^{\theta}\right]$ and  parametrization in the local basis given by Eq. ($\ref{B_r_theta}$). In Fig. \ref{fig:EVPA_contour}b we see that for predominantly radial magnetic fields the maximum EVPA reached in the image takes a low value leading to an  approximately vertical polarization vector around the image. An exception is configurations with a radial magnetic field and plasma velocity direction close to $\chi=0$ and $\chi=-\pi$. For such configuration the EVPA spans its full range along the image. When the polar component of the magnetic field increases, the maximum EVPA reached in the image increases as well. However, its variation around the image remains very limited being confined to a narrow interval below the maximum value. Thus, the polarization vector takes approximately a constant direction along the image which is strongly correlated with the magnitude of the polar component of the magnetic field. For purely polar magnetic field the maximum EVPA reaches its upper limit $|\text{max EVPA}| = \pi/2$ with variation along the image less that $1\%$. This behavior is illustrated in Fig. $\ref{fig:EVPA_image}$ where we demonstrate the evolution of the EVPA distribution for three consecutive magnetic field configurations and a constant plasma velocity direction $\chi=-\pi/2$.

Finally, we consider a magnetic field in the $(\hat\phi,\hat\theta)$ plane $\vec{B}=\left[0,\hat{B}^{\phi},\hat{B}^{\theta}\right]$ with parametrization in the local basis according to Eq. ($\ref{B_phi_theta}$). We observe a similar behavior as in the previous case. For predominantly azimuthal magnetic fields the maximum value of the EVPA is low and the polarization vector is approximately vertical along the image. This behavior is universal for any plasma velocity direction. As the polar component of the magnetic field  grows the maximum value of the EVPA increases but the variation of the EVPA stays confined to a narrow interval. This leads to an almost uniform direction of the EVPA along the image as the tilt of the polarization vector is proportional to the magnitude of the polar component of the magnetic field. For polar magnetic fields the maximum EVPA reaches its upper limit and its variation is still very small. The polarization vector is almost uniformly horizontal along the image. Examples of this behavior are presented in Fig. $\ref{fig:EVPA_image}$.

These studies demonstrate that the polar component of the magnetic field has significant impact on the EVPA around the inner direct images. Except for some isolated configurations with a predominantly radial magnetic field and an almost tangential plasma velocity, the magnitude of polar magnetic component strongly correlates with the polarization vector direction. Moreover, this direction stays almost constant around the inner image. Due to the described correlation this effect may be used as a probe for the presence of non-equatorial magnetic fields in such regular geometries.

\section{Conclusion}

We study the polarization properties of regular compact objects motivated by effective theories of quantum gravity. We consider static spherically symmetric spacetimes which generalize the Schwarzschild black hole by regularizing the central singularity. The regularization is controlled by a single length scale parameter resulting in spacetimes with de Sitter cores such as the Bardeen, Hayward and Frolov geometries or with a Minkowski core such as the Simpson-Visser geometry. These spacetimes describe  regular black holes for low value of the regularization parameters which transform continuously into horizonless particle-like configurations after certain critical value is reached.

Our analysis of the polarization behavior is based on a simple phenomenological model of a thin magnetized fluid ring which orbits in the equatorial plane and emits synchrotron radiation. We construct the polarized images of the circular emitter and analyse the observable intensity of the linearly polarized radiation and the twist of the polarization direction around the image. We concentrate on direct images since they are most relevant for the observations and examine the polarization properties in comparison with the Schwarzschild black hole for different inclinations and an asymptotic observer.

Our studies demonstrate that at low inclinations the polarization pattern  deviates very weakly from the Schwarzschild black hole both for regular black holes and particle-like geometries. This creates a challenge to distinguish them by means of polarimetric observations as potential compact objects at the galactic targets M87* and Sgr A*. 

On the other hand, at high inclinations the particle-like geometries lead to a clear-cut observational signature. A second direct image of the emitting fluid ring is produced on the observer's sky which is absent for black holes. Depending on the properties of the plasma medium the polarization intensity on this additional image may grow with three orders of magnitude providing a significant observational feature. The intensity of polarized emission is regulated by the magnetic field and velocity directions of the emitting fluid. It is amplified  by predominantly polar or axial magnetic fields and a radially inward fluid velocity. On the other hand, a predominantly radial magnetic field and a radially outward fluid velocity inhibit the polarized emission. Since in a realistic accretion model the magnetic field direction and the fluid velocity are not expected to be constant but vary with time, this will lead to flare-ups in the polarized intensity of the additional image. The presence of such fluctuations may be regarded as another signature for detecting such a class of particle-like compact objects. 

We further observed that the direction of the polarization vector around the second image is controlled by the magnetic field. It is strongly correlated with the magnitude of the polar magnetic component and for magnetic fields with a small radial component it preserves an almost constant direction around the image. The presence of such a correlation may be used as a probe for non-equatorial magnetic fields in similar regular geometries.

\clearpage
\section*{Appendix}
\subsection*{A. Comparison of the observable polarization with the Schwarz\-schild black hole}

\begin{figure}[t!]
    \centering
    \includegraphics[width=14.2cm]{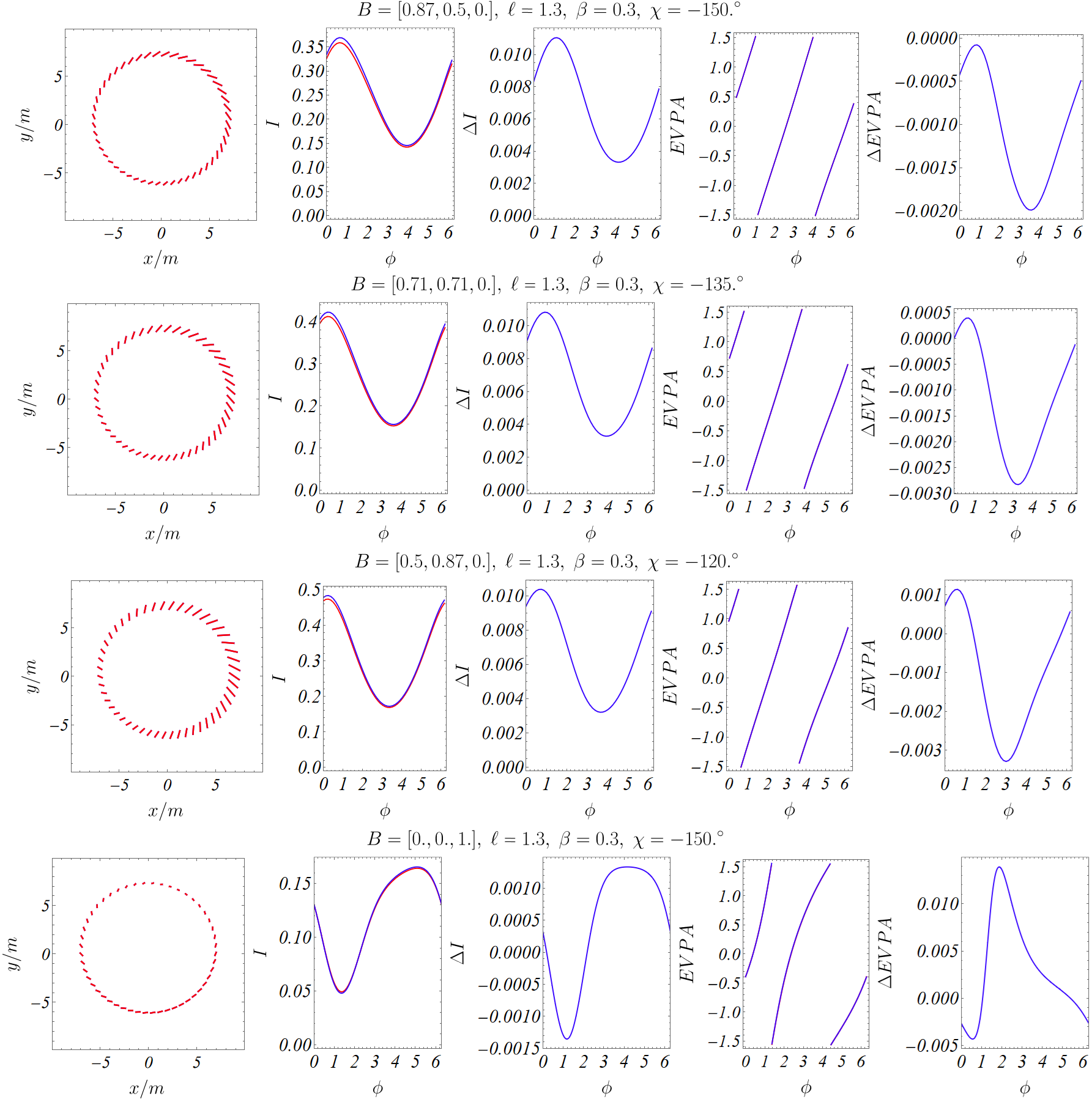}
    \caption{\small{Linear polarization for the Hayward geometry with $\ell=1.3$ (blue), compared with the Schwarzschild reference (red), for an observer inclination $i=20^\circ$. The Schwarzschild emitting orbit is located at $r_{\rm source}=6m$, and the observer at $r_{\rm obs}=10^4m$. The panels show the polarized intensity $I$, the EVPA, and the corresponding differences $\Delta I$ and $\Delta{\rm EVPA}$ relative to the Schwarzschild case.}}
    \label{fig:Hayward_vs_Shwarzschild_i=20_l=1.3}
\end{figure}

We analyze further the polarization properties of the regular geometries in comparison with the Schwarzschild black hole in order to investigate the influence of the observer's inclination. For the purpose we select a higher inclination angle which still belongs to the region in the parametric space where the circular orbits possess a single image on the observer's sky (see Fig. $\ref{fig:Hayward_vs_Shwarzschild_i=20_l=1.3}$ and Fig. $\ref{fig:Bardeen_Frolov_inner_image}$). This ensures that the regular geometries do not lead to new observational signatures so they can be distinguished from the Schwarzschild black hole only by means of the quantitative deviation in the polarization intensity and EVPA.

\begin{table}[h!]
    \centering
      \begin{tabular}{|c|c|c|c|c|}
       \hline
       \thead{ $i = 70^{\circ}$ }& \multicolumn{2}{c|}{{$B = [0.87, 0.5, 0]$, $\chi=-150\degree$}} & \multicolumn{2}{c|}{{$B = [0.71, 0.71, 0]$, $\chi=-135\degree$}} 
          \\  \hline
          
         \thead{ $\ell$}   &  \thead{$\left(\frac{\text{max}\,\Delta \text{I}}{\text{I}_\text{Sch}} \,, \, \phi \, \right)$} & \thead{$\left(\frac{\text{max}\,\Delta \text{EVPA}}{\text{EVPA}_\text{Sch}} \,, \, \phi \, \right)$} &  \thead{$\left(\frac{\text{max}\,\Delta \text{I}}{\text{I}_\text{Sch}} \,, \, \phi \, \right)$} & \thead{$\left(\frac{\text{max}\,\Delta \text{EVPA}}{\text{EVPA}_\text{Sch}} \, , \, \phi \, \right)$} 
          \\  \hline
                 \multicolumn{5}{|c|} {Hayward}
          \\  \hline
           \thead{$\vspace{0.1mm}\text{$0.75$}$\vspace{0.1mm}}  &  \thead{(0.0303, 0.51$\pi$)} & \thead{(0.0026, 0.31$\pi$)} &  \thead{(0.0265, 0.45$\pi$)} & \thead{(0.0045, 0.25$\pi$)} 
          \\  \hline

          \thead{\vspace{0.1mm}$\text{$0.85$}$\vspace{0.1mm}} &  \thead{(0.0388, 0.51$\pi$)} & \thead{(0.0034, 0.31$\pi$)} &  \thead{(0.0340, 0.45$\pi$)} & \thead{(0.0058, 0.25$\pi$)} 
          \\  \hline

          \thead{\vspace{0.1mm}$\text{$1.3$}$\vspace{0.1mm}}  &  \thead{(0.0903,  0.52$\pi$)} & \thead{(0.0079, 0.31$\pi$)} &  \thead{(0.0796,0.46$\pi$)} & \thead{(0.0134, 0.25$\pi$)} 
          \\  \hline
                  \multicolumn{5}{|c|} {Simpson}
          \\  \hline
          
           \thead{$\vspace{0.1mm}\text{$0.70$}$\vspace{0.1mm}}  &  \thead{(0.1816, 0.05$\pi$)} & \thead{(0.0417, 0.30$\pi$)} &  \thead{(0.4377, 0.46$\pi$)} & \thead{(0.0704, 0.24$\pi$)} 
          \\  \hline

          \thead{\vspace{0.1mm}$\text{$0.77$}$\vspace{0.1mm}} &  \thead{(0.1983, 0.05$\pi$)} & \thead{(0.0460, 0.30$\pi$)} &  \thead{(0.4834, 0.47$\pi$)} & \thead{(0.0782, 0.24$\pi$)} 
          \\  \hline

          \thead{\vspace{0.1mm}$\text{$0.87$}$\vspace{0.1mm}}  &  \thead{(0.2218, 0.05$\pi$)} & \thead{(0.0521, 0.30$\pi$)} &  \thead{(0.5476, 0.48$\pi$)} & \thead{(0.0882, 0.24$\pi$)} 
          \\  \hline
              \multicolumn{5}{|c|} {Bardeen}
          \\  \hline
          
           \thead{$\vspace{0.1mm}\text{$0.45$}$\vspace{0.1mm}}  &  \thead{(0.1119, 0.51$\pi$)} & \thead{(0.0095, 0.31$\pi$)} &  \thead{(0.0985, 0.45$\pi$)} & \thead{(0.0163, 0.25$\pi$)} 
          \\  \hline

          \thead{\vspace{0.1mm}$\text{$0.55$}$\vspace{0.1mm}} &  \thead{(0.1469, 0.52$\pi$)} & \thead{(0.0124, 0.31$\pi$)} &  \thead{(0.1275, 0.45$\pi$)} & \thead{(0.0212, 0.25$\pi$)} 
          \\  \hline

          \thead{\vspace{0.1mm}$\text{$0.7$}$\vspace{0.1mm}}  &  \thead{(0.2026, 0.53$\pi$)} & \thead{(0.0171, 0.31$\pi$)} &  \thead{(0.1735, 0.45$\pi$)} & \thead{(0.0290, 0.25$\pi$)} 
          \\  \hline
                            \multicolumn{5}{|c|} {Frolov}
          \\  \hline
          
           \thead{\vspace{0.1mm}$\text{$0.63$}$\vspace{0.1mm}} &  \thead{(0.0279, 0.51$\pi$)} & \thead{(0.0024, 0.31$\pi$)} &  \thead{(0.0245, 0.45$\pi$)} & \thead{(0.0043, 0.26$\pi$)} 
          \\  \hline

          \thead{\vspace{0.1mm}$\text{$0.73$}$\vspace{0.1mm}} &  \thead{(0.0390, 0.51$\pi$)} & \thead{(0.0034, 0.31$\pi$)} &  \thead{(0.0343, 0.45$\pi$)} & \thead{(0.0059, 0.26$\pi$)} 
          \\  \hline

          \thead{\vspace{0.1mm}$\text{$1.05$}$\vspace{0.1mm}} &  \thead{(0.0899, 0.52$\pi$)} & \thead{(0.0078, 0.31$\pi$)} &  \thead{(0.0785, 0.46$\pi$)} & \thead{(0.0134, 0.25$\pi$)} 
          \\  \hline

       \end{tabular}
      \caption{\small{Maximum deviations of the polarized intensity and EVPA from the Schwarzschild reference for an observer inclination $i=70^\circ$ and the
magnetic field configurations specified in the table. For each regular geometry and value of $\ell$, we report $\max(\Delta I)/I_{\rm Sch}$ and $\max(\Delta{\rm EVPA})/{\rm EVPA}_{\rm Sch}$, together with the azimuthal angle $\phi$ at which each maximum occurs. The Schwarzschild reference corresponds to an emitting orbit at $r_{\rm source}=6m$, with
the observer located at $r_{\rm obs}=10^4m$. The three values of $\ell$ for each solution represent from top to bottom a black-hole configuration, an ultracompact configuration, and a horizonless configuration without photon spheres.}}
    \label{table:theta_70_1}
\end{table}

\begin{table}[h!]
    \centering
      \begin{tabular}{|c|c|c|c|c|}
       \hline
       \thead{ $i = 70^{\circ}$ }& \multicolumn{2}{c|}{{$B = [0.5, 0.87, 0]$, $\chi=-120\degree$}} & \multicolumn{2}{c|}{{$B = [0, 0, 1]$, $\chi=-150\degree$}} 
          \\  \hline
          
         \thead{ $\ell$}   &  \thead{$\left(\frac{\text{max}\,\Delta \text{I}}{\text{I}_\text{Sch}} \,, \, \phi \, \right)$} & \thead{$\left(\frac{\text{max}\,\Delta \text{EVPA}}{\text{EVPA}_\text{Sch}} \,, \, \phi \, \right)$} &  \thead{$\left(\frac{\text{max}\,\Delta \text{I}}{\text{I}_\text{Sch}} \,, \, \phi \, \right)$} & \thead{$\left(\frac{\text{max}\,\Delta \text{EVPA}}{\text{EVPA}_\text{Sch}} \, , \, \phi \, \right)$} 
          \\  \hline
          \multicolumn{5}{|c|} {Hayward}
          \\  \hline
           \thead{$\vspace{0.1mm}\text{$0.75$}$\vspace{0.1mm}}  &  \thead{(0.0223, 0.40$\pi$)} & \thead{(0.0074, 0.20$\pi$)} &  \thead{(0.0140, 0.11$\pi$)} & \thead{(0.0921, 0.49$\pi$)} 
          \\  \hline

          \thead{\vspace{0.1mm}$\text{$0.85$}$\vspace{0.1mm}} &  \thead{(0.0285, 0.40$\pi$)} & \thead{(0.0094, 0.20$\pi$)} &  \thead{(0.0180, 0.11$\pi$)} & \thead{(0.1159, 0.49$\pi$)} 
          \\  \hline

          \thead{\vspace{0.1mm}$\text{$1.3$}$\vspace{0.1mm}}  &  \thead{(0.0655,  0.40$\pi$)} & \thead{(0.0217, 0.20$\pi$)} &  \thead{(0.0417, 0.11$\pi$)} & \thead{(0.2412, 0.49$\pi$)} 
          \\  \hline
                  \multicolumn{5}{|c|} {Simpson}
          \\  \hline
          
           \thead{$\vspace{0.1mm}\text{$0.70$}$\vspace{0.1mm}}  &  \thead{(0.3555, 0.39$\pi$)} & \thead{(0.1039, 0.18$\pi$)} &  \thead{(0.2695, 0.10$\pi$)} & \thead{(0.5421, 0.39$\pi$)} 
          \\  \hline

          \thead{\vspace{0.1mm}$\text{$0.77$}$\vspace{0.1mm}} &  \thead{(0.3892, 0.39$\pi$)} & \thead{(0.1138, 0.18$\pi$)} &  \thead{(0.2961, 0.10$\pi$)} & \thead{(0.5671, 0.39$\pi$)} 
          \\  \hline

          \thead{\vspace{0.1mm}$\text{$0.87$}$\vspace{0.1mm}}  &  \thead{(0.4368, 0.39$\pi$)} & \thead{(0.1274, 0.18$\pi$)} &  \thead{(0.3342, 0.10$\pi$)} & \thead{(0.6131, 0.38$\pi$)} 
          \\  \hline
                         \multicolumn{5}{|c|} {Bardeen}
          \\  \hline
          
           \thead{$\vspace{0.1mm}\text{$0.45$}$\vspace{0.1mm}}  &  \thead{(0.0817, 0.39$\pi$)} & \thead{(0.0259, 0.19$\pi$)} &  \thead{(0.0574, 0.11$\pi$)} & \thead{(0.2781, 0.49$\pi$)} 
          \\  \hline

          \thead{\vspace{0.1mm}$\text{$0.55$}$\vspace{0.1mm}} &  \thead{(0.1058, 0.39$\pi$)} & \thead{(0.0337, 0.19$\pi$)} &  \thead{(0.0747, 0.11$\pi$)} & \thead{(0.3402, 0.49$\pi$)} 
          \\  \hline

          \thead{\vspace{0.1mm}$\text{$0.7$}$\vspace{0.1mm}}  &  \thead{(0.1439, 0.39$\pi$)} & \thead{(0.0459, 0.19$\pi$)} &  \thead{(0.1024, 0.11$\pi$)} & \thead{(0.4733, 0.50$\pi$)} 
          \\  \hline
                            \multicolumn{5}{|c|} {Frolov}
          \\  \hline
          
           \thead{\vspace{0.1mm}$\text{$0.63$}$\vspace{0.1mm}} &  \thead{(0.0205, 0.40$\pi$)} & \thead{(0.0069, 0.20$\pi$)} &  \thead{(0.0135, 0.12$\pi$)} & \thead{(0.0800, 0.48$\pi$)} 
          \\  \hline

          \thead{\vspace{0.1mm}$\text{$0.73$}$\vspace{0.1mm}} &  \thead{(0.0285, 0.40$\pi$)} & \thead{(0.0096, 0.20$\pi$)} &  \thead{(0.0188, 0.12$\pi$)} & \thead{(0.1083, 0.48$\pi$)} 
          \\  \hline

          \thead{\vspace{0.1mm}$\text{$1.05$}$\vspace{0.1mm}} &  \thead{(0.0651, 0.40$\pi$)} & \thead{(0.0216, 0.20$\pi$)} &  \thead{(0.0430, 0.12$\pi$)} & \thead{(0.2343, 0.49$\pi$)} 
          \\  \hline

       \end{tabular}
      \caption{\small{Maximum deviations of the polarized intensity and EVPA from the Schwarzschild reference for an observer inclination $i=70^\circ$ and the
magnetic field configurations specified in the table. For each regular geometry and value of $\ell$, we report $\max(\Delta I)/I_{\rm Sch}$ and $\max(\Delta{\rm EVPA})/{\rm EVPA}_{\rm Sch}$, together with the azimuthal angle $\phi$ at which each maximum occurs. The Schwarzschild reference corresponds to an emitting orbit at $r_{\rm source}=6m$, with the observer located at $r_{\rm obs}=10^4m$. The three values of $\ell$ for each solution represent from top to bottom a black-hole configuration, an ultracompact configuration, and a horizonless configuration without photon spheres.}}
    \label{table:theta_70_2}
\end{table}

\begin{figure}[t!]
    \centering
    \includegraphics[width=14.2cm]{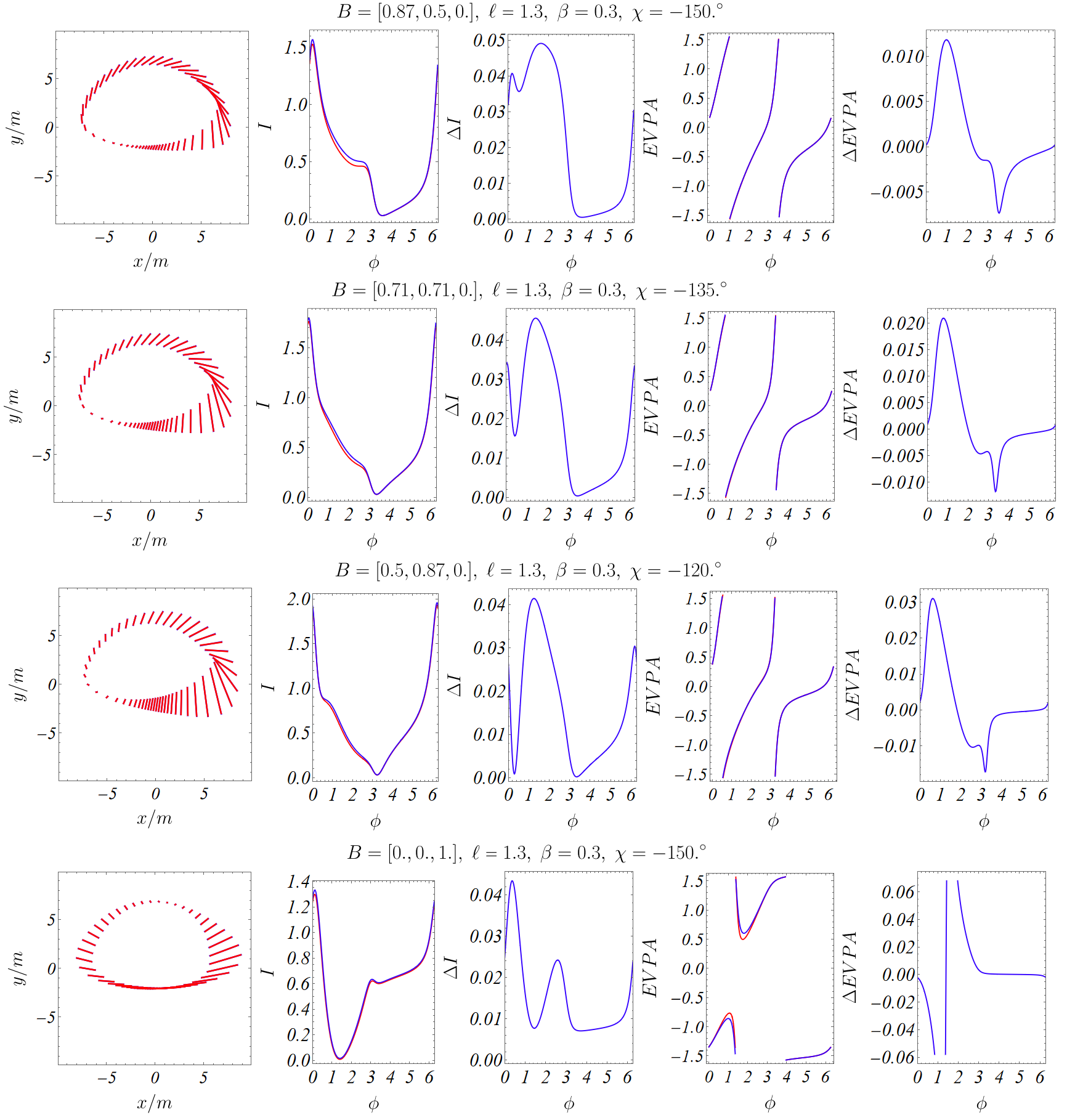}
    \caption{\small{Linear polarization for the Hayward geometry with $\ell=1.3$ (blue), compared with the Schwarzschild reference (red), for an observer inclination $i=70^\circ$. The Schwarzschild emitting orbit is located at $r_{\rm source}=6m$, and the observer at $r_{\rm obs}=10^4m$. The panels show the polarized intensity $I$, the EVPA, and the corresponding differences $\Delta I$ and $\Delta{\rm EVPA}$ relative to the Schwarzschild case.}}
    \label{fig:Hayward_vs_Shwarzschild_i=70_l=1.3}
\end{figure}

\begin{figure}[t!]
    \centering
    \includegraphics[width=14.2cm]{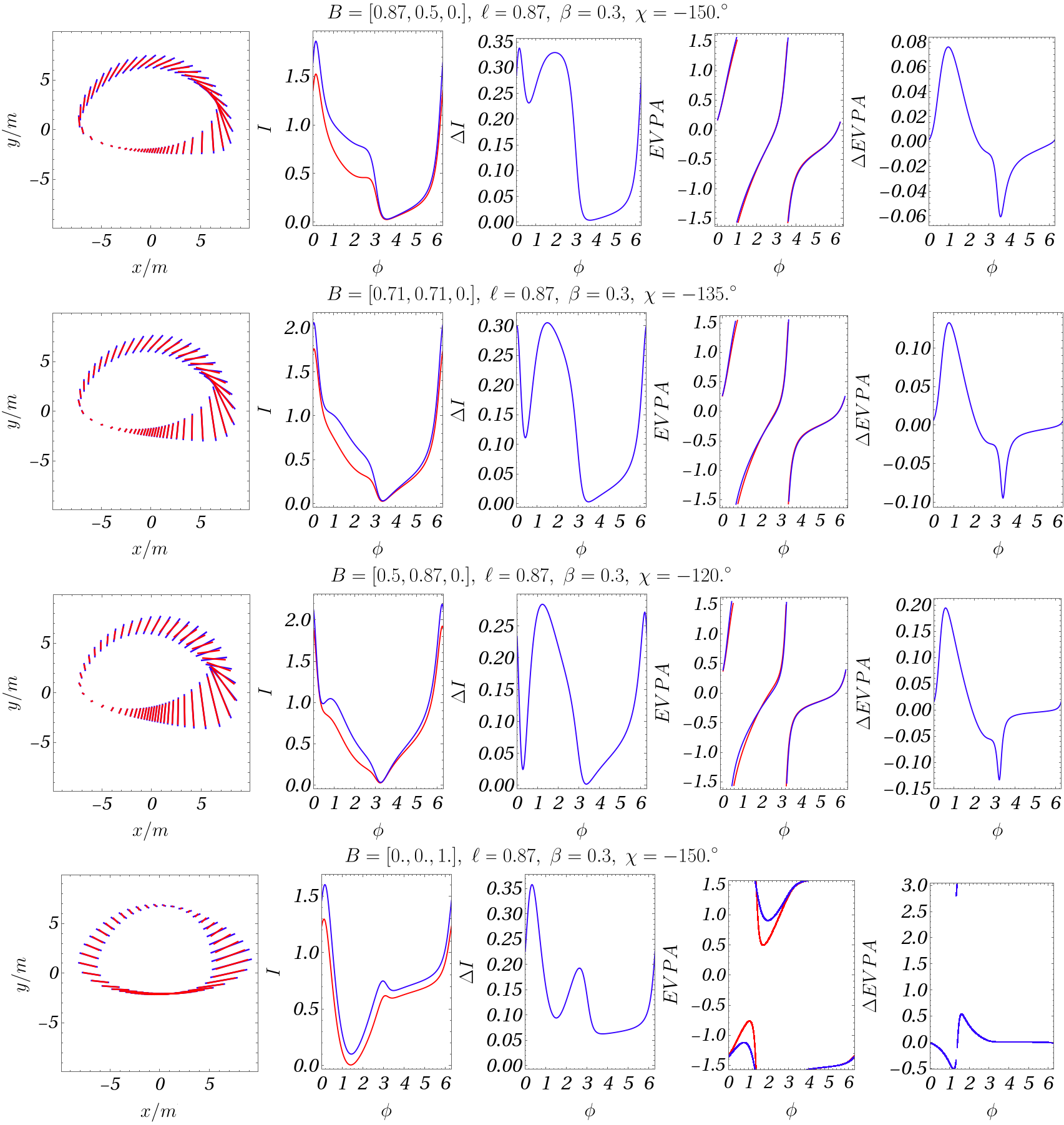}
    \caption{\small{Linear polarization for the Simpson geometry with $\ell=0.87$ (blue), compared with the Schwarzschild reference (red), for an observer inclination $i=70^\circ$. The Schwarzschild emitting orbit is located at $r_{\rm source}=6m$, and the observer at $r_{\rm obs}=10^4m$. The panels show the polarized intensity $I$, the EVPA, and the corresponding differences $\Delta I$ and $\Delta{\rm EVPA}$ relative to the Schwarzschild case.}}
    \label{fig:Simpson_vs_Shwarzschild_i=70_l=0.87}
\end{figure}

In Tables $\ref{table:theta_70_1}$-$\ref{table:theta_70_2}$ we present our analysis of the deviation of the polarization properties of the four regular geometries from the Schwarzschild black hole for observer's inclination angle $i=70^{\circ}$. We follow the investigations in Section 5 evaluating the maximum deviation in the polarization intensity and EVPA along the circular orbit image for several magnetic field configurations and representative values for the regularization parameter $\ell$. The deviation increases substantially in comparison with the lower inclination $i=20^{\circ}$ reaching a maximum value of $\approx 50\%$ for the polarization intensity and $\approx 61\%$ for the EVPA among all the configurations. We observe similar correlations as in our previous study for lower inclination. Regular geometries with de Sitter core lead to lower deviations than the Simpson spacetime with Minkowski core. The deviation grows with increasing the regularization parameter $\ell$ showing most pronounced differences from the Schwarzschild black hole for particle-like geometries without light rings. For all the regular geometries  the deviation in the polarization intensity is largest for equatorial magnetic fields with a substantial radial component reaching a maximal value of $\approx 55\%$ for the Simpson particle-like solution and $\approx 9\% - 20\%$ for the particle-like solutions with de Sitter core. On the other hand, for all the regular geometries the deviation in the EVPA is largest for vertical  magnetic field configurations reaching $\approx 61\%$ for the Simpson particle-like solution and $\approx 23\%-47\%$ for the particle-like solutions with de Sitter cores. In Figs. $\ref{fig:Hayward_vs_Shwarzschild_i=70_l=1.3}$- $\ref{fig:Simpson_vs_Shwarzschild_i=70_l=0.87}$ we provide an  illustration the polarization properties for higher inclination angles in comparison with the Schwarzschild black hole showing the Hayward and Simpson particle-like solutions without light rings as representative examples.

\subsection*{B. Formation of inner direct images}

\begin{figure}[t!]
    \centering
    \includegraphics[width=14.2cm]{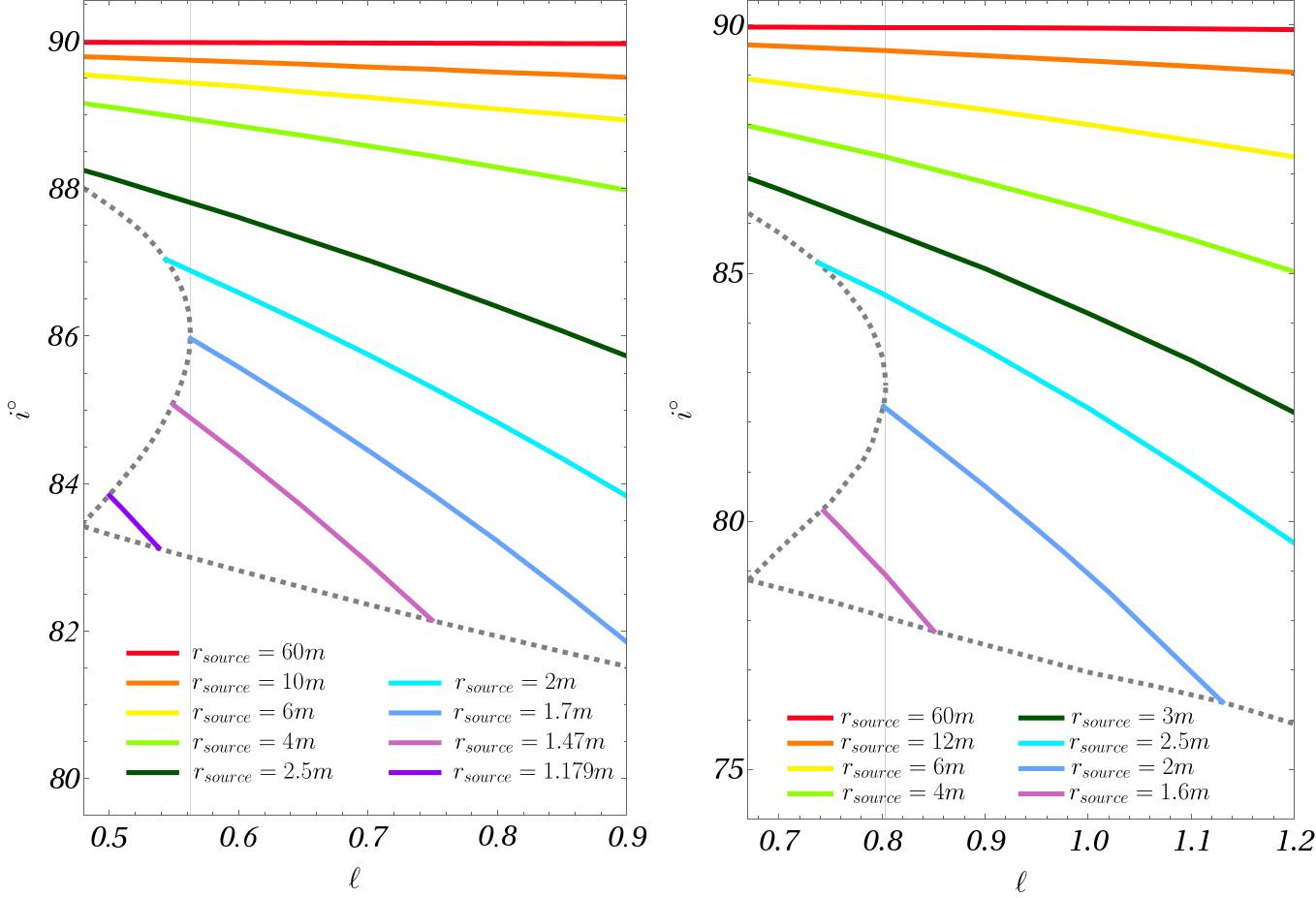}
    \caption{{\small Parameter space for the existence of the inner direct images in the in the Bardeen (left) and Frolov (right) spacetimes. The vertical axis gives the critical inclination for the formation of inner images, while the horizontal axis represents the regularization parameter $\ell$. The different curves correspond to different radial positions of the emitting orbit $r_{source}$ with color coding given by the legend.  As $r_{source}$ increases, the curves approach a horizontal asymptote which can be approximated with the red curve. We denote by dashed lines the location of the stable and unstable photon spheres and the location of the ICOs. The vertical grey line corresponds to the regularization parameter $\ell_{ph}$ where the two photon spheres merge and provides the limit between ultracompact geometries and geometries without light rings. The observer is located at $r_{obs}=10^{4}\,m$.}}
    \label{fig:Bardeen_Frolov_inner_image}
\end{figure}

We study the parameter space for the formation of the inner direct images for Bardeen and Frolov particle-like solutions as a supplementary analysis to the investigations of the Hayward and Simpson spacetimes performed in Section 6 (see Fig. $\ref{fig:Hayward_Simpson_inner_image}$). The parameter space is presented in Fig. $\ref{fig:Bardeen_Frolov_inner_image}$ where we illustrate the critical inclination necessary to observe inner direct images as a function of the regularization parameter $\ell$ and the emitting radius $r_{source}$. We observe similar correlations as for the Hayward and Simpson geometries. The critical inclination increases with the radial distance of the emitting orbit and for larger distances it depends weakly on the regularization parameter. Circular orbits with $r_{source} \gtrsim 6m$ lead to an inner direct image only for nearly edge-on inclinations of the accretion disk. For lower emitting radii $r_{source} \lesssim 3m$ the inner direct images becomes observable after a lower critical inclination which further declines with growing regularization parameter $\ell$. The critical inclination cannot decrease arbitrary but it is limited due to the existence of ICOs and restrictions on the natural values of  $\ell$. Adopting a scale for the regularization parameter $\ell\sim 1$ we obtain that emitting sources with $r_{source} \lesssim 3m$ produce inner direct images for regularization parameters $82^{\circ} \lesssim i \lesssim 87^{\circ}$ in
the Bardeen spacetime and  for $76^{\circ} \lesssim i \lesssim 87^{\circ}$ in the Frolov spacetime. Thus, the behavior of the parameter space in the Frolov spacetime closely resembles the Hayward geometry while the Bardeen spacetime requires significantly larger inclinations for inner direct images to become observable.

\section*{Acknowledgments}
We gratefully acknowledges support from the Bulgarian National Science Fund (NSF) under Grant KP-06-DV/8 within the funding programme “VIHREN–2024”.

\end{document}